\documentclass{aa}
\usepackage{txfonts}
\usepackage[pdfpagelabels=false,colorlinks=true,linkcolor=blue,citecolor=blue,filecolor=blue,urlcolor=blue]{hyperref}

\usepackage[utf8]{inputenc}
\usepackage{amsmath}
\usepackage{graphicx}
\usepackage{float}
\usepackage{placeins}

\usepackage{color}
\usepackage{xcolor} 
\usepackage{soul}
\newcommand{\adr}[1]{\textcolor{black}{#1}}

\newcommand{\equ}[1]{eq.~(\ref{eq:#1})}

\newcommand{\se}[1]{\S\ref{sec:#1}}
\newcommand{\fig}[1]{Fig.~\ref{fig:#1}}

\newcommand{\Fig}[1]{Figure~\ref{fig:#1}}

\newcommand{\be}{\begin{equation}}
\newcommand{\ee}{\end{equation}}
\newcommand{\ba}{\begin{align}}
\newcommand{\ea}{\end{align}}
\newcommand{\bad}{\begin{equation} \begin{aligned}}
\newcommand{\ead}{\end{aligned} \end{equation}}
\newcommand{\bea}{\begin{eqnarray}}
\newcommand{\eea}{\end{eqnarray}}
\def\ra{\rangle}
\def\la{\langle}

\newcommand{\bul}{$\bullet\ $}

\newcommand{\no}{\noindent}

\newcommand{\msun}{M_\odot}
\newcommand{\Msun}{M_\odot}

\newcommand{\ifm}[1]{\relax\ifmmode#1\else$\mathsurround=0pt #1$\fi}
\newcommand{\kms}{\ifmmode\,{\rm km}\,{\rm s}^{-1}\else km$\,$s$^{-1}$\fi}

\newcommand{\kpc}{\,{\rm kpc}}
\newcommand{\pc}{\,{\rm pc}}
\newcommand{\cm}{\,{\rm cm}}
\newcommand{\Gyr}{\,{\rm Gyr}}

\newcommand{\Myr}{\,{\rm Myr}}
\newcommand{\myr}{\,{\rm Myr}}

\newcommand{\cmc}{\,{\rm cm}^{-3}}

\newcommand{\ltsima}{$\; \buildrel < \over \sim \;$}
\newcommand{\lsim}{\lower.5ex\hbox{\ltsima}}
\newcommand{\gtsima}{$\; \buildrel > \over \sim \;$}
\newcommand{\gsim}{\lower.5ex\hbox{\gtsima}}
\newcommand{\prop}{\propto}
\newcommand{\dd}{{\rm d}}
\newcommand{\pa}{\partial}

\def\Mv{M_{\rm v}}

\def\Mveight{M_{{\rm v},10.8}}

\def\Rv{R_{\rm v}}

\def\Ms{M_{\rm s}}

\def\Re{R_{\rm e}}

\def\Sig1{\Sigma_1}

\def\Md{M_{\rm disc}}
\def\Rd{R_{\rm disc}}
\def\Hd{H_{\rm disc}}

\def\Mc{M_{\rm c}}

\def\Vc{V_{\rm c}}
\def\Vrot{V_{\rm rot}}

\def\td{t_{\rm d}}

\def\fb{f_{\rm b}}

\def\Mbh{M_{\rm bh}}

\def\eps2{\epsilon_{-2}}

\def\trlx{t_{\rm rlx}}

\def\eps{\epsilon}
\def\tcc{t_{\rm cc}}

\def\ssim{\!\sim\!}
\def\seq{\!=\!}
\def\ssimeq{\!\simeq\!}
\def\sequiv{\!\equiv\!}
\def\sgt{\!>\!}
\def\slt{\!<\!}
\def\sgsim{\!\gsim\!}
\def\slsim{\!\lsim\!}
\def\sgeq{\!\geq\!}
\def\sleq{\!\leq\!}
\def\sgg{\!\gg\!}
\def\sll{\!\ll\!}
\def\sdash{\!-\!}
\def\stimes{\!\times\!}
\def\sprop{\!\propto\!}
\def\spm{\!\pm\!}
\def\splus{\!+\!}

\def\MT{M_{\rm T}}

\def\sigr{\sigma_r}

\def\Vd{V_{\rm d}}

\def\mmax{m_{\rm max}}

\def\zffb{z_{\rm ffb}}

\def\fbh{f_{\rm bh}}
\def\tdf{t_{\rm df}}
\def\adf{a_{\rm df}}

\def\Vd{V_{\rm d}}
\def\Vh{V_{\rm h}}
\def\Md{M_{\rm d}}
\def\Mh{M_{\rm h}}
\def\mmax{m_{\rm max}}
\def\Rot{{\cal R}}

\def\Vk{V_{\rm recoil}}

\def\Vexp{V_1}

\def\rexp{r_1}

\def\rhalt{r_{\rm halt}}

\def\rmax{r_{\rm max}}
\def\Vmax{V_{\rm max}}
\def\Vesc{V_{\rm esc}}

\def\Vrel{V_{\rm rel}}
\def\Fd{F_{\rm d}}

\def\Rhill{R_{\rm H}}

\defcitealias{dekel23}{D23}

\begin{document}

\titlerunning{Black holes in the FFB Scenario}
\title{Growth of Massive Black-Holes in FFB Galaxies at Cosmic Dawn}

\authorrunning{Dekel et al.}
\author{
Avishai Dekel$^{1,2}$ \fnmsep\thanks{E-mail: \href{mailto:dekel@huji.ac.il}{dekel@huji.ac.il}},
Nicholas C. Stone$^{1}$,
Dhruba Dutta Chowdhury$^{1}$,
Shmuel Gilbaum$^{1}$,
Zhaozhou Li$^1$,\\ 
Nir Mandelker$^1$,
\and
Frank C. van den Bosch$^{3}$
}

\institute{
$^1$Racah Institute of Physics, The Hebrew University, Jerusalem 91904 Israel\\
$^2$SCIPP, University of California, Santa Cruz, CA 95064, USA\\
$^3$Department of Astronomy, Yale University, New Haven, CT 06520, USA
}

\AANum{A\&A}

\large  





\abstract
{}
{The scenario of feedback-free starbursts (FFB), which predicts excessively 
bright galaxies at cosmic dawn as observed using JWST, may provide a natural 
setting for black hole (BH) growth.  This involves the formation of 
intermediate-mass seed BHs and their runaway mergers into super-massive BHs 
with high BH-to-stellar mass ratios and low AGN luminosities.}
{We present a scenario of merger-driven BH growth in FFB galaxies
and study its feasibility.}
{BH seeds form within the building blocks of the FFB galaxies, 
namely, thousands of compact star clusters, each starbursting in a free-fall 
time of a few Myr 
before the onset of stellar and supernova feedback.
The BH seeds form by rapid core collapse in the FFB clusters,
in a few free-fall times, 
sped up by the migration of massive stars due to the young, 
broad stellar mass function and stimulated by a `gravo-gyro' 
instability due to internal cluster rotation and flattening.
BHs of $\sim\!10^4\msun$ are expected in $\sim\!10^6\msun$ FFB clusters 
within sub-kpc galactic disks at $z\ssim 10$.
The BHs then migrate to the galaxy center by dynamical friction, 
hastened by the compact FFB stellar galactic disk configuration.
Efficient mergers of the BH seeds will produce 
$\sim\!10^{6\sdash 8}\msun$ BHs with 
a BH-to-stellar mass ratio $\sim\! 0.01$ by $z \ssim 4\sdash 7$, as observed.
The growth of the central BH by mergers can overcome the bottleneck 
introduced by gravitational wave recoils if the BHs inspiral within a 
relatively cold disk or if the escape velocity from the galaxy is boosted by 
a wet compaction event.
Such events, common in massive galaxies at high redshifts,
can also help by speeding up the inward BH migration and
by providing central gas to assist with the final parsec problem.}
{The cold disk version of the FFB scenario provides a feasible 
route for the formation of supermassive BHs.}


\maketitle

\section{Introduction}
\label{sec:intro}

The largest supermassive black holes (SMBHs) observed at high redshifts have 
long posed a challenge for simple models of SMBH formation and evolution.  
The small population of observed high-$z$ SMBHs with mass $\gtrsim\! 10^9\msun$ 
\citep{barth03, mortlock11, banados16} are too large to have grown from 
Eddington-limited accretion onto stellar-mass black hole (BH) seeds 
\citep{haiman01}, suggesting that some exotic process must quickly assemble 
intermediate-mass BH seeds in the early Universe \citep{volonteri10}.  
The three most commonly discussed scenarios are, in increasing order of seed 
mass, (i) Pop III stellar remnants \citep{schneider02}; 
(ii) runaway collisions in dense star clusters \citep{portegies04}; 
and (iii) direct collapse of low-metallicity gas clouds \citep{loeb94}.  
At present, it is not clear which if any of these scenarios is responsible 
for the most extreme high-$z$ SMBHs \citep{inayoshi20}.

More recently, a second puzzle has emerged concerning the high-$z$ SMBH
formation and evolution. 
JWST observations at $z \seq 4\sdash 7$
from the CEERS and JADES surveys \citep{harikane23,maiolino23,ubler23}
indicate SMBHs of masses in the range $10^{7\pm 1}\msun$
with especially high BH-to-stellar mass ratios, $\fbh$, in the ball park of 
$\fbh \ssim 0.01$. 
A statistical analysis by \citet{pacucci23}, based on 
a relatively uniform subsample of 21 massive BHs and their host galaxies,  
yields a bias-corrected BH-to-stellar mass relation that is 10 to 100 times
above the standard relation at low redshifts \citep[e.g.][]{reines15}, 
indicating a deviation larger than $3\sigma$ between the two.
While an alternative statistical analysis of these data by 
\citet{li_silverman24}
cautions that the difference between the relations at high and low 
redshifts may be less dramatic, the evidence for a high $\fbh$ at high redshifts
is intriguing and it calls for a theoretical understanding.

We investigate here how such SMBHs may naturally arise within the
scenario of {\it feedback-free starbursts} (FFB) \citep{dekel23,li24},
which predicts excessively bright galaxies at cosmic dawn as observed using
JWST. 
This physical model envisions high star-formation efficiency in free-fall 
collapses of thousands of dense gas clouds within compact galaxies, 
and thus explains the JWST-observed excess of bright galaxies at $z \ssim 10$
compared to the standard theory that is valid at later epochs, where 
stellar feedback suppresses star formation.
At first glance, the extreme efficiency of FFB star formation may seem to be
in competition with the need for efficient BH formation and a high $\fbh$,
but we argue that the FFB scenario can actually provide a natural setting
for the SMBH challenge. 
Our current analysis is meant as a feasibility study, aimed at verifying the
conditions under which FFB galaxy formation at cosmic dawn can enable the 
formation of SMBHs with high $\fbh$ ratio by $z \seq 4\sdash 7$,
and to what extent the SMBH growth can be driven by BH mergers.

We address four crucial stages in the BH growth process.
First, the formation of intermediate-mass seed BHs by sped-up `core collapse'
\citep{lynden68} within the FFB star clusters at cosmic dawn.
Second, the subsequent inward migration of these BHs within the  
FFB disk galaxies by dynamical friction 
\citep[][DF]{chandrasekhar43}
against the stars and dark matter in the compact galaxies.
Third, overcoming a bottleneck in the initial growth of the SMBHs 
subject to gravitational-wave (GW) recoils after BH mergers
\citep{pretorius05,campanelli06,baker06}.
Fourth, the effect of `wet compaction' events \citep{zolotov15,lapiner23}
on avoiding possible dynamical-friction stalling  
at the end of the second stage's DF-driven inspiral
\citep{read06,kaur18,banik21},
preventing escape by recoils in the third stage \citep{madau04,blecha08}, 
and helping with viscous last parsec approach \citep{begelman80}.

The building blocks of an FFB galaxy are thousands of young star clusters, 
each formed in a feedback-free starburst during a free-fall time of a few Myr.
Such clusters are expected to develop core collapse driven by two-body
star-star relaxation, and may under certain conditions form massive central 
stellar objects that soon after collapse to seed BHs 
\citep{spitzer71,portegies04,devecchi09,katz15,rantala24}.
In order to produce massive seed BHs with a high $\fbh$,
the core-collapse should occur in less than $\sim\!3 \Myr$, 
the characteristic lifetime of the massive stars \citep{hirschi07}.
During that period, the core collapse is enhanced by the mass segregation
associated with the presence of the massive stars and the BH growth is not 
suppressed by supernova and stellar feedback.
The core collapse can be sped up by three special features of
the FFB clusters, namely, their compactness, the presence of 
young massive stars in them, and their internal rotation and spatial 
flattening, as follows.

Compactness is the basic feature of the FFB clusters at cosmic dawn.
They are expected to have an internal 3D density of
$n \ssim n_{\rm crit} \ssim 3\stimes 10^{3} \cmc$, which allows free-fall
collapse before the onset of feedback as well as cooling below $10^4$K
during the free fall \citep{dekel23}.
The clusters are also expected to have a surface density of 
$\Sigma \ssim \Sigma_{\rm crit} \ssim 3\times 10^3\msun\pc^{-2}$
such that they are gravitationally confined against outflows by radiative
feedback \citep{menon23,grudic24}. 

Core collapse is sped up by mass segregation, 
the inward migration of young massive stars, 
which are naturally present during the 
feedback-free period of $\sim\!3\Myr$.   
The high-$z$ initial stellar mass function (IMF) is rather uncertain
and could differ from the standard IMF at lower redshifts.
In particular, it may be top-heavy, 
possibly made of pop-III stars \citep{bromm02}, 
or a later population as obtained in certain simulations 
\citep{grudic24,bate23} 
and as argued based on certain observations \citep{cameron23,steinhardt23},
though these interpretations are controversial \citep{tacchella24}.
A top-heavy IMF, with an enhanced UV luminosity-to-stellar mass ratio
of a few to ten, will also add to the excessive brightness as observed by JWST
\citep{yung24} while it will not suppress the feedback-free high star formation
efficiency \citep{menon24}.
Such a top-heavy IMF, with a larger fraction of massive stars,
should help speed up the core collapse.

Star-star interactions would be boosted if the clusters are spatially flattened 
and the core collapse would be sped up further when the clusters are 
rotating, giving rise to a `gravo-gyro' instability \citep{hachisu79}.  
This would be natural in the disk version of the FFB scenario \citep{dekel23}, 
where the star-forming clouds, of $\sim\!10^7\msun$ and below,
fragment from Toomre-unstable rotating gaseous galactic disks
\citep{toomre64,dsc09}. 
Indeed, clumps formed in this way are expected to be
largely supported by rotation and to be flattened correspondingly 
\citep{ceverino12}.

After estimating the core-collapse time and the resultant BH
seed mass as a function of the cluster mass and radius, its flattening and 
rotation support, and the IMF in it, we will evaluate the range of FFB 
clump properties that permit core collapse as rapid as 3 Myr. 
This will allow us to estimate the population of
seed BHs produced during the FFB phase at cosmic dawn, 
and predict the corresponding $\fbh$ in these clusters.
We estimate below that BH seeds of $\sim \!10^4 \msun$ can typically form 
in $\sim\!10^6\msun$ clusters within the FFB disks at $z\ssim 10$. 

Following the $z \ssim 10$ FFB phase of a galaxy, 
the system of thousands of star clusters with their central seed BHs is  
expected to evolve within the compact galactic disk of radius 
$\sim\!200\pc$ \citep[][and references therein]{li24}. 
The clusters of internal velocities $\sim\!10\kms$, orbiting in a potential well
of $\sim\!100\kms$, are expected to tidally disrupt each other in a few 
orbital times to form a centrally concentrated galactic stellar disk 
within a common dark-matter halo. 
Subsequently, the seed BHs (and any surviving clusters) are expected to 
migrate toward the galaxy center by dynamical friction, where they could
coalesce into a SMBH.
In order to produce massive SMBHs with high values of $\fbh$ by 
$z\ssim 4\sdash 7$, as observed,
most of the seed BHs should migrate to the center in less than a Gigayear.
Such an efficient dynamical friction would require large enough BH seed masses,
and it would benefit from the compact, high density galactic FFB configuration.
The DF is expected to be particularly efficient in the disk version of the FFB 
scenario, where the relative velocities of the orbiting BHs and the 
surrounding stars are low, especially if the profiles of 
density and angular velocity are steeply declining with radius to avoid
DF core stalling.
We will evaluate the expected migration timescales in FFB galaxies and verify
the conditions for the required efficient inward migration.

A bottleneck in the initial growth of the SMBH by BH mergers could be caused
by the GW recoils of the merger remnants 
\citep[e.g., as simulated by][]{sijacki09}.
For major
mergers of mass ratios between 0.1 and 0.8, the recoil velocity, as computed
by numerical general relativity, could be larger than the central escape 
velocity from the galaxy, especially in cases of BH spin-orbit misalignments.
The FFB galaxies, being both massive and compact, are expected to
provide relatively high, confining escape velocities.
However, the spin-orbit alignments necessary for non-ejective recoils
would require that the seed BHs be inspiraling through a rather cold galactic
disk configuration. 
We will quantify the conditions for overcoming this recoil bottleneck.

SMBH growth in the last two stages can be boosted by events 
of wet compaction into baryon-dominated `nuggets'.
As seen in simulations \citep{zolotov15,lapiner23,degraaff24}
and as observed \citep{barro13,dokkum15,barro17,huertas18,degraaff24},
this process is generic in the high-$z$ history of galaxies,
preferably when the DM halo exceeds a `golden mass' of $\sim\!10^{11.5}\Msun$
\citep{dekel19_gold}.
It has far-reaching implications on all major galaxy properties
\citep{tacchella16_prof,tacchella16_ms,tomassetti16,lapiner23}.
Triggered by drastic angular-momentum loss, e.g.,
due to wet galaxy mergers or collisions of counter-rotating inflowing streams,
gas is pushed to the central regions of the galaxies.
This results in a cuspy gas-rich `blue nugget' that passively evolves to a
compact stellar `red nugget', allowing the formation of a stable, extended
gaseous disk/ring around it \citep{dekel20_ring}.
Such compaction events may assist SMBH growth by 
(i) enhancing DF migration of the BH seeds, which is of special importance 
when the original galaxy structure may cause `core stalling',
(ii) increasing the escape velocity of the host galaxy, helping to retain BH 
merger products against GW recoil, and
(iii) providing central gas to help solving the final parsec problem.

%
%
%

The outline of this article is as follows.
In \se{seeds} we estimate the seed BH population formed by core collapse
in the FFB star clusters at cosmic dawn.
In \se{DF} we evaluate the inward migration of the BHs and the expected 
SMBH mass and $\fbh$ at $z \ssim 4 \sdash 7$.
In \se{recoil} we evaluate the potential effect of GW recoil
of the SMBH at the early phases of SMBH growth. 
In \se{compaction} we address the possible assistance to SMBH growth by
compaction events.
Finally, in \se{conc} we summarize our conclusions.

\section{Observed BH-Stellar Mass Relation}
\label{sec:obs}

\begin{figure} 
\centering
\includegraphics[width=0.49\textwidth] 
{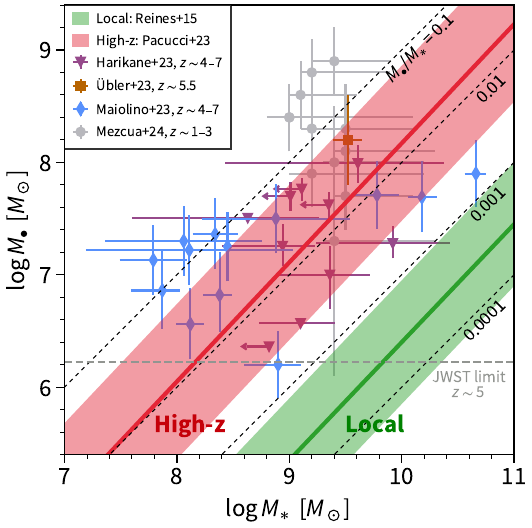}
\vspace{-15pt}
\caption{
Black-hole mass versus stellar mass as estimated from observations.
Shown is the bias-corrected relation at $z\seq 4\sdash 7$
as derived by \citet{pacucci23} (red shading, $\pm 1\sigma$ scatter).
It is based on the data from JWST marked by the symbols
\citep[JADES, CEERS and GA-NIFS][]{maiolino23,harikane23,ubler23}.
Also shown are data at $z \seq 1 \sdash 3$ \citep{mezcua24}.
The high-$z$ relation is compared to the local relation by \citet{reines15} 
(green shading).
The typical values of $\fbh$ at $z \seq 4\sdash 7$ are higher than the local 
relation by $>\!3\sigma$, a factor of $10\sdash 100$.
}
\vspace{-10pt}
\label{fig:obs}
\end{figure}

Black-hole masses and their host stellar masses have been estimated 
for galaxies with low-luminosity AGN from the CEERS, JADES and GA-NIFS 
surveys using JWST in the 
redshift range $z \seq 4 \sdash 7$ \citep{harikane23,maiolino23,ubler23}. 
These black holes are likely only the tip of the iceberg,
and the derived BH-to-stellar mass relation has to be bias-corrected for
the coupled effects of selection biases and the $\sim\!1$dex measurement
uncertainties in both the BH and stellar masses.
\citet{pacucci23} performed a statistical analysis of a relatively
uniform sample of 21 AGN from these surveys of massive BHs,
eight from \citet{harikane23}, twelve from \citet{maiolino23} and one from 
\citet{ubler23}.
These are all spectroscopically confirmed with NIRSpec, their black hole
masses are estimated with the H$\alpha$ line, 
their host stellar masses are estimated by UV to optical SED fitting,
and lensed objects or double AGN candidates were not included.
They obtained at $z \seq 4 \sdash 7$ the bias-corrected relation
\be
\log \left(\frac{\Mbh}{M_\odot}\right) = -2.43 (\pm 0.83) + 1.06 (\pm 0.09)
\log \left(\frac{\Ms}{M_\odot}\right) \, ,
\label{eq:BS}
\ee
with a scatter of 0.69 dex.
In comparison, the standard local relation, obtained
by \citet{reines15} using more massive galaxies at low redshifts,  
is of a similar functional form with the numerical coefficients 
$-4.10 (\pm 0.08)$ and $1.05 (\pm 0.11)$ respectively.
\citet{pacucci23} estimate a deviation larger than $3\sigma$ between these 
relations, with the values of $\fbh$ at $z \seq 4\sdash 7$ 
on average higher than the local
relation by a factor of $10$ to $100$.
The data points and the bias-corrected $\fbh$ ratios are shown in \fig{obs},
based on \citet{pacucci23}, with additional data by \citet{mezcua24} 
at $z \seq 1\sdash 3$.
The origin of this discrepancy is a major open issue.

We note in passing that even more extreme cases are detected, 
at higher redshift and/or with a very high $\fbh$, 
such as UHZ1 at $z \seq 10.1$ with $\fbh$ of order 
unity \citep{bogdan24,kovacs24,maiolino24,natarajan24}.
In the current feasibility study we focus on the origin of the average 
properties of the main body of massive BHs at $z \ssim 4-7$, and comment 
on the possible origin of scatter about the mean relation.

\section{Black-Hole Seeds in FFB Star Clusters}
\label{sec:seeds}

\adr{
Given the efficient star formation in FFB star clusters at cosmic dawn,
one might suspect that it would come at the expense of BH growth in these
clusters. We show here that, quite the contrary, the FFB clusters are likely
to provide an ideal setting for the formation of intermediate-mass black 
holes (IMBH), and attempt to characterize the population of IMBHs  
that are expected to form by core collapse in the FFB clusters.
}

\subsection{Core collapse in a young cluster with a broad IMF}

Stellar clusters are expected to suffer core collapse due to two-body
relaxation that induces gravo-thermal instability \citep{lynden68}. 
In this case, 
energy transfer from a kinematically hot gravitating core to the 
cooler outer envelope makes the core contract and heat up further
due to the virial theorem (or the associated negative heat capacity of 
gravitating systems), leading to a runaway process. This leads to 
a very massive central star (VMS)
\citep{portegies99} which collapses to a black hole that
may contain as much as one percent of the cluster mass
\citep{yungelson08,portegies02,heger03,rantala24,fujii24}.
In a cluster younger than a few Myr, when the massive stars are 
still alive on the main sequence, 
the core collapse is significantly sped up by the inward migration of the 
massive stars due to mass segregation.
Thus, 
in order to efficiently form a massive seed BH at the center of the cluster,
the core collapse should occur in less than 
$\sim \! 3\Myr$, the main-sequence lifetime of massive stars of 
$\sim\!40\msun$ \citep{hirschi07}. 
This time is safely before the onset of feedback \citep[][Fig.~1]{dekel23}
that otherwise might have suppressed the VMS/BH growth.  
The rapid inward migration of the massive stars, 
and their disappearance into the VMS/BH while they are still on the main
sequence, prevents them from ever generating effective stellar feedback.
Thus, the formation of a massive seed BH is intimately related to the FFB 
phase in the cluster, either as a cause or as an effect.
The important quantities to be evaluated are the time for core collapse 
and the resultant black-hole mass.

Too comments are worth mentioning in passing.
First,
if the massive stars in the clusters indeed coalesce
with the central BHs before they explode as SNe, the feedback from post-FFB
clusters would be somewhat weakened, contributed by lower-mass stars.
This may enhance the shielding of new FFB clusters beyond the original
estimate of \citet{dekel23}, who conservatively concluded that the clouds have 
to be more massive than $10^4\msun$ for survival. 
This could also lower even further
the gas fraction, metallicity, dust content and outflow strength in FFB
galaxies as evaluated in \citet{li24}.
On the other hand, encounters between main sequence stars and BHs may produce
micro-tidal disruption events \citep{perets16}, which can generate 
`accretion feedback' in the form of radiative luminosity or mechanical 
luminosity from the ensuing period of super-Eddington accretion driving
outflows.

Second, we comment that high values of $\fbh$ in massive BHs were 
predicted as a result of direct gas collapse in nuclear star clusters (NSCs),
in a series of papers from \citet{lodato06} to \citet{alexander14}
and \citet{natarajan24}. In this different picture, the first massive star 
that collapses into a BH grows initially via wind-fed accretion as it random 
walks through the star cluster colliding with the stars.  
The situation in this NSC is essentially free of feedback, partially resembling
the FFB conditions that we propose are valid in the thousands of FFB clusters 
in a massive galaxy at cosmic dawn.

\subsubsection{Core collapse time}

In a cluster of mass $\Mc$ and half-mass radius $\Re$,
if it consists of $N$ equal-mass stars, as can be approximated 
in old globular clusters, 
the two-body relaxation time is 
\be 
\trlx = \frac{0.138\,N}{\ln(\gamma N)}
\left( \frac{\Re^3}{G\,\Mc} \right)^{1/2} \, ,
\label{eq:trlx}
\ee
where $\gamma \ssim 0.11$ \citep{giersz94}.
The core-collapse time in this case is rather long, 
$\tcc \ssim 15\sdash20\, \trlx$.
In a multi-mass system, on the other hand, the core collapse occurs on the
shorter segregation time,
due to energy equipartition or dynamical friction.
Based on \citet{spitzer71} and \citet{portegies04},
when a standard \citet{kroupa01} IMF is assumed,
the segregation time can be evaluated by a similar expression to \equ{trlx}
but with $N$ replaced by $\Mc/\mmax$ (rather than $N \seq \Mc/\la m \ra$), 
where $\mmax$ is the mass of the most massive star in the cluster.
This gives a core-collapse time of $\tcc \ssim 0.2\,\trlx$, which is
\be
\tcc = 1.31 \Myr\,
\left( \frac{\Re}{1\pc} \right)^{3/2}
\left( \frac{\Mc}{10^{5}\msun} \right)^{1/2}
\left( \frac{\mmax}{100\msun} \right)^{-1}
\left( \frac{\ln \Lambda}{4.7} \right)^{-1} \, .
\label{eq:tcc_rizzuto}
\ee
Here $\Lambda \ssimeq \gamma\,\Mc/\mmax$, which gives  
$\ln\Lambda \ssim 4.7$ for $\gamma \seq 0.11$. 
%
This estimate has been crudely confirmed by simulations 
\citep{rizzuto21,rantala24}.
It is in the same ball park as the alternative estimate
by \citet{portegies02} and \citet{devecchi09} where 
the IMF is characterized by $\la m \ra$ rather $\mmax$,
\be
\tcc \simeq 3 \Myr\, 
\left( \frac{\Re}{1\pc} \right)^{3/2}
\left( \frac{\Mc}{10^{5.7}\msun} \right)^{1/2}
\left( \frac{\la m \ra}{10\msun} \right)^{-1}
\left( \frac{\ln\Lambda_{\rm c}}{8.5} \right)^{-1} \, .
\label{eq:tcc_port}
\ee
Here the Coulomb logarithm $\ln\Lambda_{\rm c} \simeq \ln(0.1\,\Mc/\la m \ra)$ 
is with respect to $8.5$,
the value assumed for $\Mc \seq 10^{5.7}\msun$ and $\la m \ra \seq 10\msun$.

\begin{figure*} 
\centering
\includegraphics[width=0.49\textwidth,trim={1.5cm 5.5cm 1.5cm 2.8cm},clip]
{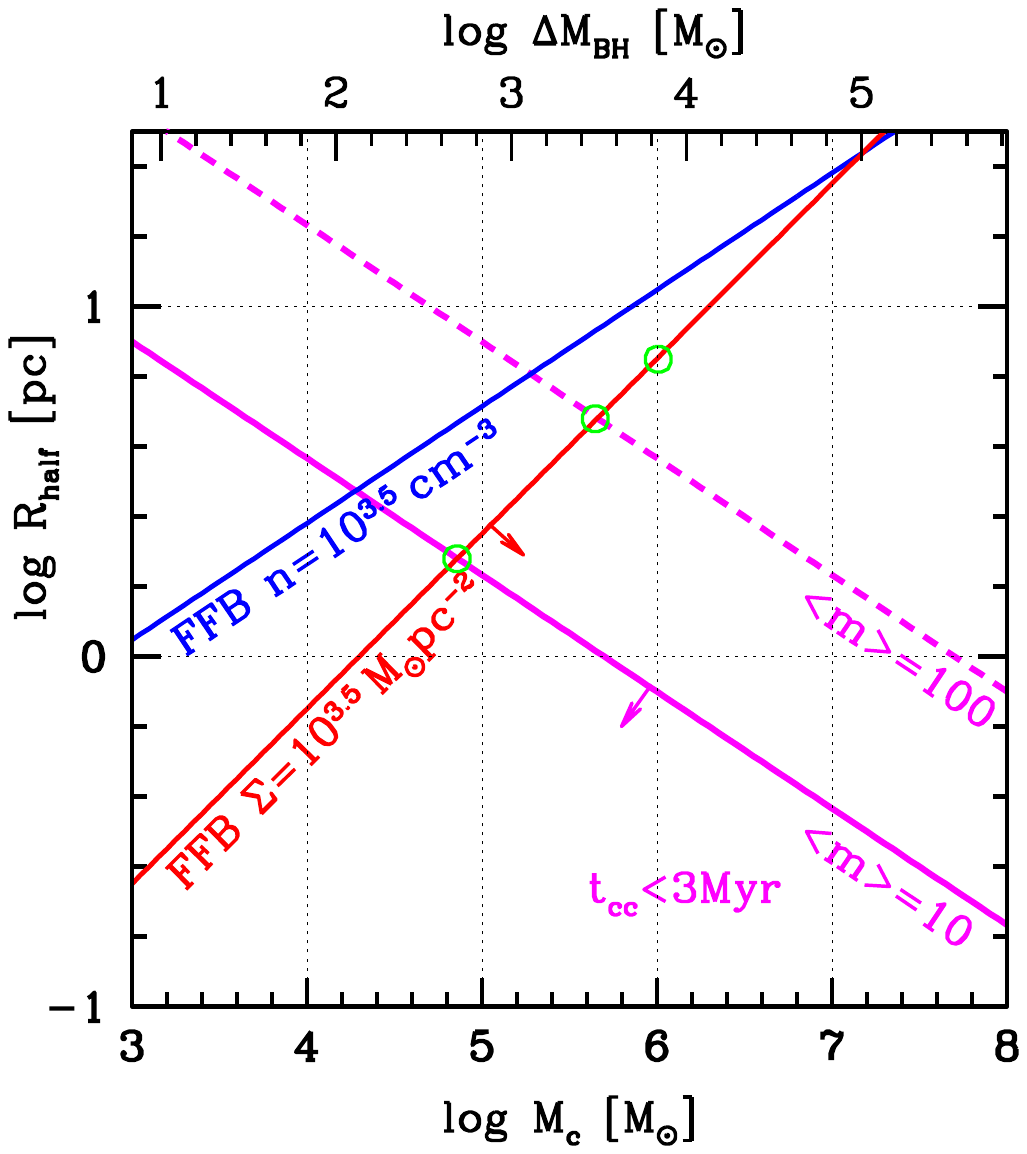}
\includegraphics[width=0.49\textwidth,trim={1.5cm 5.5cm 1.5cm 2.8cm},clip]
{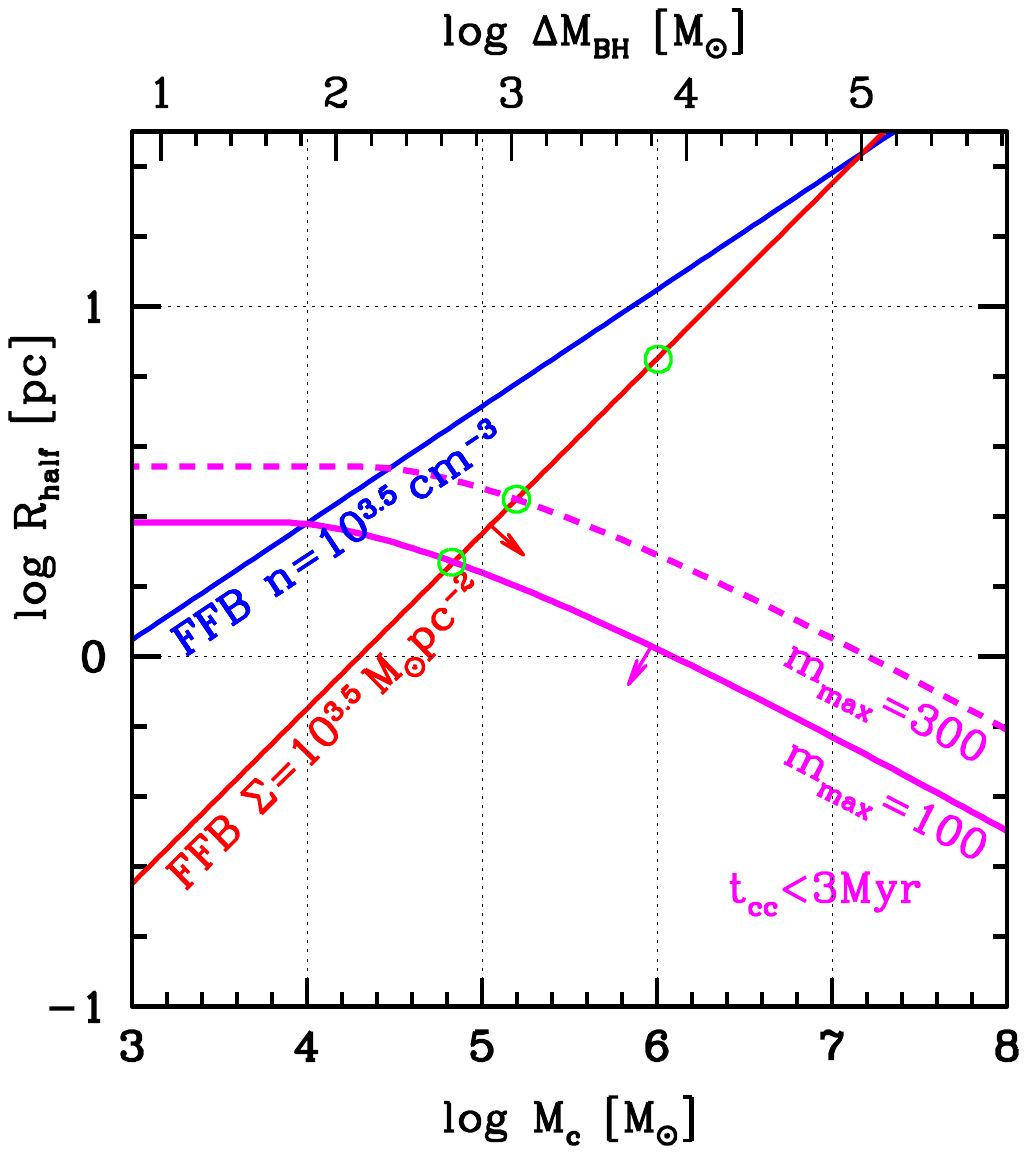}
\vspace{-5pt}
\caption{
Black hole growth in non-rotating FFB clusters with a broad IMF.
Shown is the cluster half-mass radius versus mass.
The magenta curves and regions below them
refer to clusters where core collapse occurs within $3\Myr$, the lifetime of
massive stars.
{\bf Left:} based on \equ{tcc_port} \citep{portegies02,devecchi09},
assuming either $\la m \ra \seq 10\msun$ (solid)
or $\la m \ra \seq 100\msun$ (dashed).
{\bf Right:} based on \equ{tcc_rizzuto} \citep[][eq.~4]{rizzuto21},
assuming $m_{\rm max} \seq 100$ (solid) or $300\msun$ (dashed).
The blue and red lines refer to constant density $n\seq 10^{3.5} \cm^{-3}$
and surface density $\Sigma \seq 10^{3.5} \msun \pc^{-2}$ within the half-mass
radius.
Feedback-free starbursts are expected to occur near such densities.
FFB clouds of the Jeans mass and radius, $\sim\! 10^6\msun$ and $\sim\! 7\pc$,
lie above the core-collapse line, though not too far away
if the IMF is very top-heavy,
of a population-III.1 type, with $\la m \ra \ssim 100\msun$ (left)
or $m_{\rm max} \sgt 300\msun$ (right).
If the sub-Jeans clouds form by fragmentation with the same $\Sigma$, they
should lie along the red line, such that core collapse will occur in $3\Myr$
for $\Mc \leq 10^5\msun$ and $\Re \leq 2\pc$.
The corresponding black-hole mass at the end of the FFB phase, \equ{mbh_port},
in addition to the initial mass of the massive star that became the VMS,
is marked in the top axis, indicating $\Mbh \ssim 10^3\msun$ in
$\sim\!10^5\msun$ clusters.
}
\vspace{-10pt}
\label{fig:bh_imf}
\end{figure*}

\subsubsection{Black Hole Mass}
\label{sec:mass}

For low metallicity, the stellar mass loss is small, so the growth of the VMS
is efficient.
Once the VMS is more massive than $260\msun$, the resultant black hole is 
expected to retain most of the mass of the VMS \citep{heger03}.
According to \citet{portegies02},
the black hole mass at $\tcc$ is expected to be
\be
\Mbh \simeq m_\star + 6.8\times 10^{-3}\, \Mc\, (\ln \Lambda_{\rm c}/{8.5}) \, ,
\label{eq:mbh_port}
\ee
where $m_\star$ is the initial mass of the massive star that became the VMS.
In the second term there should be a multiplicative factor of order unity 
which depends on $\tcc$ with respect to $3\Myr$.
For a cluster of $\Mc \ssim 10^6\msun$, this gives $\Mbh \ssim 10^4\msun$.
The high seed BH mass is a key for obtaining a high $\fbh$, 
in the ball park of 0.01, as indicated by the observations.

The black holes formed by core collapse will grow further, over longer 
timescales, due to the tidal capture and/or disruption of stars 
\citep{stone17,rizzuto23}, and possibly gas accretion as well 
\citep{schleicher22}.  Assuming that the long-term intermediate-mass BH growth
occurs only after any intra-cluster gas has been depleted or expelled due to
stellar feedback, 
we focus on the the timescale for growth by the first possibility of star
capture.  
Early stages of growth via star capture can occur in the ``full loss cone'' 
limit, but once the BH grows modestly, it will generally find itself in the 
``empty loss cone'' regime \citep{cohn78}, where the rate of stellar 
consumption is limited, and set by the rate at which two-body scatterings 
diffuse stars onto highly radial orbits that can interact strongly with the 
BH at pericenter \citep{stone17}.  Taking the approximate rate of empty loss 
cone growth to be $\dot{N}$ from Eq. 36 of \citet{stone17}, and assuming that
all clusters have the same surface mass density $\Sigma$, we find that the mass
doubling time for the BH, 
$t_{2} \seq M_{\rm bh} / (\langle m_{\rm s} \rangle \dot{N})$, 
is roughly
\be
t_{2} \ssimeq 160\Myr 
\left( \frac{M_{\rm bh}}{680\msun} \right)^{23/12}\!\! 
\left( \frac{\Mc}{10^5\msun} \right)^{-7/8}\!\! 
\left( \frac{\Sigma }{10^{3.5}\msun\!\pc^{-2}} \right)^{-7/8}\!\! .
\label{eq:doubling}
\ee
In evaluating $\dot{N}$, we have taken the average stellar mass and radius 
to be $\langle m_{\rm s} \rangle \seq 0.3 \Msun$ and 
$r_{\rm s} \seq 0.38 R_\odot$, 
respectively. We have also assumed that the cluster half-mass radius is 
$\Re \seq \Mc^{1/2} / \Sigma^{1/2}$, and that the second moment of the 
stellar mass function is $\langle m_{\rm s}^2 \rangle \seq 1\Msun^2$.  
Logarithmic factors in Eq. 36 of \citet{stone17} have been evaluated assuming 
the fiducial values for $M_{\rm bh}$, $\Mc$, and $\Sigma$ that normalize 
\equ{doubling} above, introducing mild (logarithmic) inaccuracies when 
extending this approximate formula to e.g. other BH masses.  
However, from this simple formula, 
combined with the BH mass as a function of cluster mass from \equ{mbh_port},
we can see that unless initial cluster masses are relatively large 
($\Mc \sgg 10^5\msun$), the initial mass of the BH 
will not %
be heavily modified by star capture within a time 
$t \slsim 100\Myr$. We have assumed here that, as with tidal disruption events,
half of the star's mass is eventually consumed by the BH \citep{rees88}.
We note also that this rate of star capture is calculated using a classic empty
loss cone formula that does not account for Brownian motion of the BH, 
an effect which is not well understood but which may become more important 
in the intermediate-mass range \citep{magorrian99}.

A population of growing BHs in FFB star clusters of the same 
surface density $\Sigma$ but different masses $\Mc$ will achieve a simple, 
power-law scaling, 
\be 
M_{\rm bh} \prop \Mc^{\nu} t^{\tau}\, , \quad  
\nu = 21/46\, , \quad \tau = 12/23 \, .
\label{eq:MbhMc}
\ee
This expression assumes that we are considering late enough times so that 
$t \sgt t_{2}$, 
such that $M_{\rm bh}(t) \ssimeq \langle m_{\rm s} \rangle \dot{N}\,t$.
It also assumes that the times are late enough so that 
$M_{\rm bh} \sll \Mc$, i.e., the BH has not yet consumed the majority of the 
star cluster, and that the clusters still survive tidal disruption.  
Assuming a cluster mass function $\phi \sprop \Mc^{-\alpha}$,
\equ{MbhMc} implies a late-time BH mass function within the clusters
\be
\phi(M_{\rm bh}) \prop M_{\rm bh}^{-\delta}\, , \quad 
\delta = (\alpha -1)/\nu +1 \, ,
\label{eq:delta}
\ee
at least for lower BH masses and as long as the clusters survive intact.
With a mass-function slope $\alpha \seq 1.8$ for the clusters
\citep{mandelker14,mandelker17},
the slope for the BH mass function could steepen by stellar capture to
values as large as $\delta \seq 2.75$.
A conservative version of this steepening will serve our purposes
when estimating the effect of GW recoil in \se{mf} below.

Adopting the core-collapse time estimate from \equ{tcc_port}, 
or alternatively from \equ{tcc_rizzuto}, 
\fig{bh_imf} shows the relation between cluster mass 
and half-mass radius for $\tcc \seq 3\Myr$.
This is in comparison to the FFB characteristic quantities within the half-mass 
radius, namely density $n\seq 10^{3.5} \cm^{-3}$ and surface density
$\Sigma \seq 10^{3.5} \msun \pc^{-2}$ \citep{dekel23}.
FFB clouds of the Jeans mass and radius, 
$\sim\! 10^6\msun$ and $\sim\! 7\pc$ \citep{dekel23},
lie above the line of core-collapse in $3\Myr$, but not too far away
if the IMF is very top-heavy, e.g.,
of a population-III.1 type, with $\la m \ra \ssim 100\msun$ in \equ{tcc_port} 
or $m_{\rm max} \sgt 300\msun$ in \equ{tcc_rizzuto}.
For a less top-heavy IMF, with $\la m \ra \ssim 10\msun$ or 
$m_{\rm max} \sgsim 100\msun$, the Jeans-mass clouds lie well above
the region for $\tcc \slt 3\myr$.  However,
if sub-Jeans clouds form by fragmentation with the same critical surface 
density $\Sigma$ that characterizes FFB, they should lie along the 
$\Sigma \seq const.$ red line, such that core collapse will occur for
$\Mc \leq 10^5\msun$ and $\Re \leq 2\pc$.
According to \equ{mbh_port}, the black-hole will form in such cluster's FFB 
phase with $\Mbh \ssim 10^3\msun$. This is in the intermediate
mass range of BHs, but on its relatively low side, possibly not massive enough 
for inspiraling to the galaxy center in less than a Gigayear, to be estimated 
in \se{DF} below.

\subsection{Rotating Clusters: Gravo-Gyro Instability and Flattening}

In the disk version of the FFB model,
the clumps formed by Toomre disk instability are expected to be largely 
supported by rotation \citep{ceverino12}, and they will be
spatially flattened accordingly.
This would speed up the core collapse in them due to 
(i) the gravo-gyro instability 
\citep{hachisu79,hachisu82,ernst07,kim08,hong13,kamlah22}, and
(ii) the flattening which induces a shorter two-body relaxation time and 
dynamical friction time.
Each of these effects could shorten the core-collapse time by a factor of a
few, providing a combined reduction in $\tcc$ by an order of magnitude.
These are described in the three following subsections.

\begin{figure} 
\centering
\includegraphics[width=0.49\textwidth,trim={1.5cm 5.5cm 1.5cm 4.5cm},clip]
{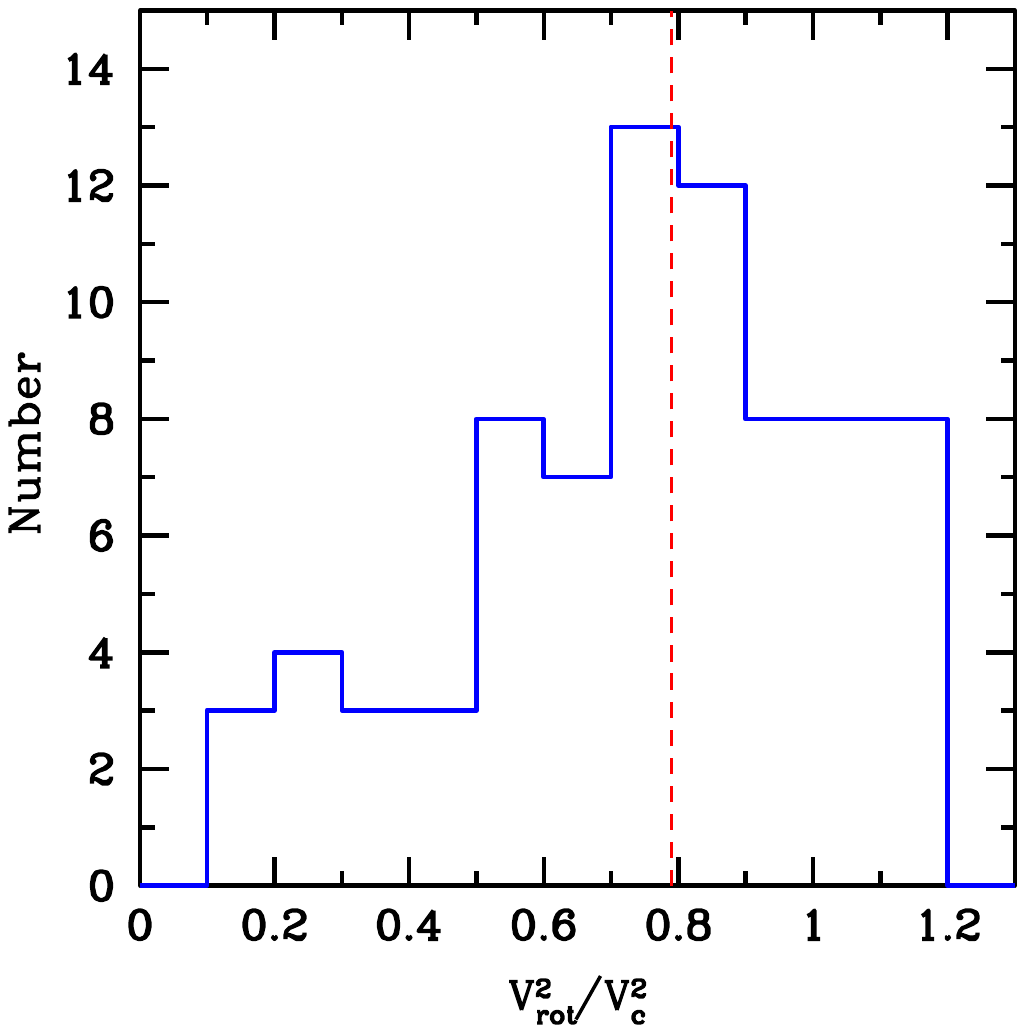}
\vspace{-15pt}
\caption{
The distribution of rotation support, ${\cal R} \seq \Vrot^2/\Vc^2$,
for clumps in high-redshift massive galactic disks,
based on zoom-in cosmological simulations \citep{ceverino12}.
Shown is the distribution over the clumps of circular velocity 
in the range $50 \sdash 250\kms$.
The median of ${\cal R} \seq 0.79$ is marked (red).
The values that exceed unity are indicative of the error that should be
assigned to the ${\cal R}$ values.
We note in comparison that the Galactic globular clusters are hardly rotating, 
typically with ${\cal R} \sll 1$ \citep{sollima19}.
}
\vspace{-10pt}
\label{fig:ceverino}
\end{figure}

\subsubsection{Rotation of clumps in simulations}

We assume that the FFB galaxies and the clouds that fragment from them at 
cosmic dawn
are qualitatively similar to the star-forming disk galaxies that are
well studied, simulated and observed, at somewhat lower redshifts near cosmic 
noon
\citep{dsc09,genzel11,ceverino12,wuyts12,mandelker14,mandelker17,guo15,guo18,
huertas20,ginzburg21,dekel22,dekel23_gclumps}.
The rotational support of clumps that form by violent disk instability in thick
gas-rich galactic disks at cosmic noon has been studied by
\citet{ceverino12} using an analytic model, zoom-in cosmological simulations,
and isolated-disk simulations.
They measured the rotation support via the quantity
$\Rot \seq \Vrot^2/\Vc^2$, where $\Vrot$ and $\Vc$ are the clump rotation and
circular velocities respectively.
Assuming Jeans equilibrium in the equatorial plane of the flattened clump, 
$\Vc^2 \seq \Vrot^2 + 2\sigr^2$, where $\sigr$ is the radial velocity
dispersion, $\Rot$ corresponds to the rotation-to-dispersion velocity ratio via
$\Vrot/\sigr \seq \sqrt{2}(\Rot^{-1}-1)^{-1/2}$.
The associated axial ratio assuming hydrostatic equilibrium perpendicular
to the major plane of rotation is $\Rd/\Hd \ssimeq \Vrot/\sigr$.

The analytic model of \citet{ceverino12},
based on Toomre disk instability and conservation of 
angular momentum during clump formation, predicts
$\Rot \ssimeq 0.2\,c$, where $c$ is the clump-radius contraction factor 
with respect to the protoclump patch, assumed to be at the disk mean density.
At Jeans equilibrium this corresponds to
$\Vrot/\sigr \simeq \sqrt{2}\,(5.1\,c^{-1}-1)^{-1/2}$.
We note that a contraction of $c\ssimeq 5$ brings the clump to full rotation
support.

The distribution of $\Rot$ over the clumps in
pre-VELA zoom-in cosmological simulations as measured in \citet{ceverino12}
is shown in \fig{ceverino}. 
The median rotation support is $\Rot \seq 0.79$.
Jeans equilibrium implies a median of $\Vrot/\sigr \seq 2.66$ for these clumps.
The most massive clumps tend to show a higher level of rotation support,
with $\Rot \geq 0.75$ spread about unity.
Jeans equilibrium is found to be valid to a good accuracy for each clump, and
$\Rot$ is indeed spread about $\Rot \seq 0.2\,c$ as predicted by the toy model,
with $c \seq 2 \sdash 8$. 
The values that exceed unity in \fig{ceverino} are indicative of the error 
that should be assigned to the ${\cal R}$ values.
They may result from clumps that deviate from equilibrium during formation or 
disruption, deviations from spherical symmetry, tidal effects from 
perturbations in the background disk, and so on. 

We note in passing that today's Galactic globular clusters, 
as observed by Gaia, are hardly rotating, 
with ${\cal R}$ values in the range $0.003 \sdash 0.14$
for the 15/62 clusters for which any rotation has been detected 
\citep[][Table 2]{sollima19}.
This distinguishes the high-redshift clusters from the local ones,
if the former originated from rotating galactic disks according to
the disk version of the FFB scenario.

\subsubsection{Gravo-Gyro instability}

A rotating cluster would develop core collapse more efficiently due to 
a combined `gravo-thermo-gyro' instability.
The more familiar gravo-thermal instability \citep{lynden68}
can be understood as a result of the negative specific heat of a
self-gravitating system, represented by the virial theorem.
When a kinematically hot core is embedded in a colder envelope,
the energy transferred out of the core causes 
the core to contract and therefore heat up towards virial equilibrium,
thus generating a runaway process.
In analogy, the gravo-gyro instability \citep{hachisu79,hachisu82}
can be interpreted as being due to the negative `specific moment of inertia'. 
When a rotating core is embedded in a slower rotating ring, AM is
transferred out of the core, causing the core to contract and thus decrease
its moment of inertia, thus making the core rotation speed up 
(the `ballerina' effect), generating a runaway process.
The two instabilities help each other, as each of them generates contraction.

Following the early simulations by \citet{ernst07} and \citet{kim08},
\citet{hong13} performed N-body simulations of a cluster, with a simplified
IMF consisting of two populations of individual stellar masses $m_2 \geq m_1$ 
and total masses $M_1 > M_2$, and they tested a range of dimensionless
angular velocities $\omega_0$, as defined in and after their Eq.~1 for a 
King model \citep{king62} following \citet{lupton87}. 
For $m_2/m_1\seq 2$ and $M_1/M_2 \seq 5$ (case M2A),
comparing $\omega_0 \seq 0$ and $1.5$, they find
$\tcc/\trlx \seq 6.8$ and $2.8$, namely a speed up by a factor of $2.4$ because
of the rotation.
For $m_2/m_1 \seq 20$ (M2D) and no rotation they obtained
$\tcc/\trlx \ssim 0.42$, namely a speed up by a factor 10 or more due to the
more top-heavy IMF. 
This implies a crude dependence on the IMF of $\tcc \sprop (m_1/m_2)$.
However, they found that with the high mass ratio the relative effect of 
rotation becomes weaker.

\citet{kamlah22} performed N-body simulations of a cluster, with a
\citet{kroupa01} IMF,
including binaries and other complications, and tested a varying level of
rotation $\omega_0\seq 0.0 \sdash 1.8$.
The cluster mass is $\Mc \seq 1.1\times 10^5\msun$, obeying a King model 
profile \citep{king62}, with parameters $W_0\seq 6.0$, 
$r_{\rm half} \seq 1.85\pc$, and $r_{\rm tidal} \seq 65.59\pc$.
For $\omega_0 \seq 1.8$ they find $\tcc \ssim 3\Myr$.
Only a minor core collapse is seen for $\omega_0 \seq 0.6$, 
at $\tcc \ssim 30\Myr$, with no clear evidence for core collapse in the 
non-rotating case.
This is based on Figs. 1 and 2 of \citet{kamlah22} for the core radius, and 
Figs. 3, 4, and 6 for the flattening.
This corresponds to a speed up by a factor of $\sim\! 10$ due to rotation.

\begin{figure} 
\centering
\includegraphics[width=0.49\textwidth,trim={1.5cm 5.5cm 1.5cm 2.8cm},clip]
{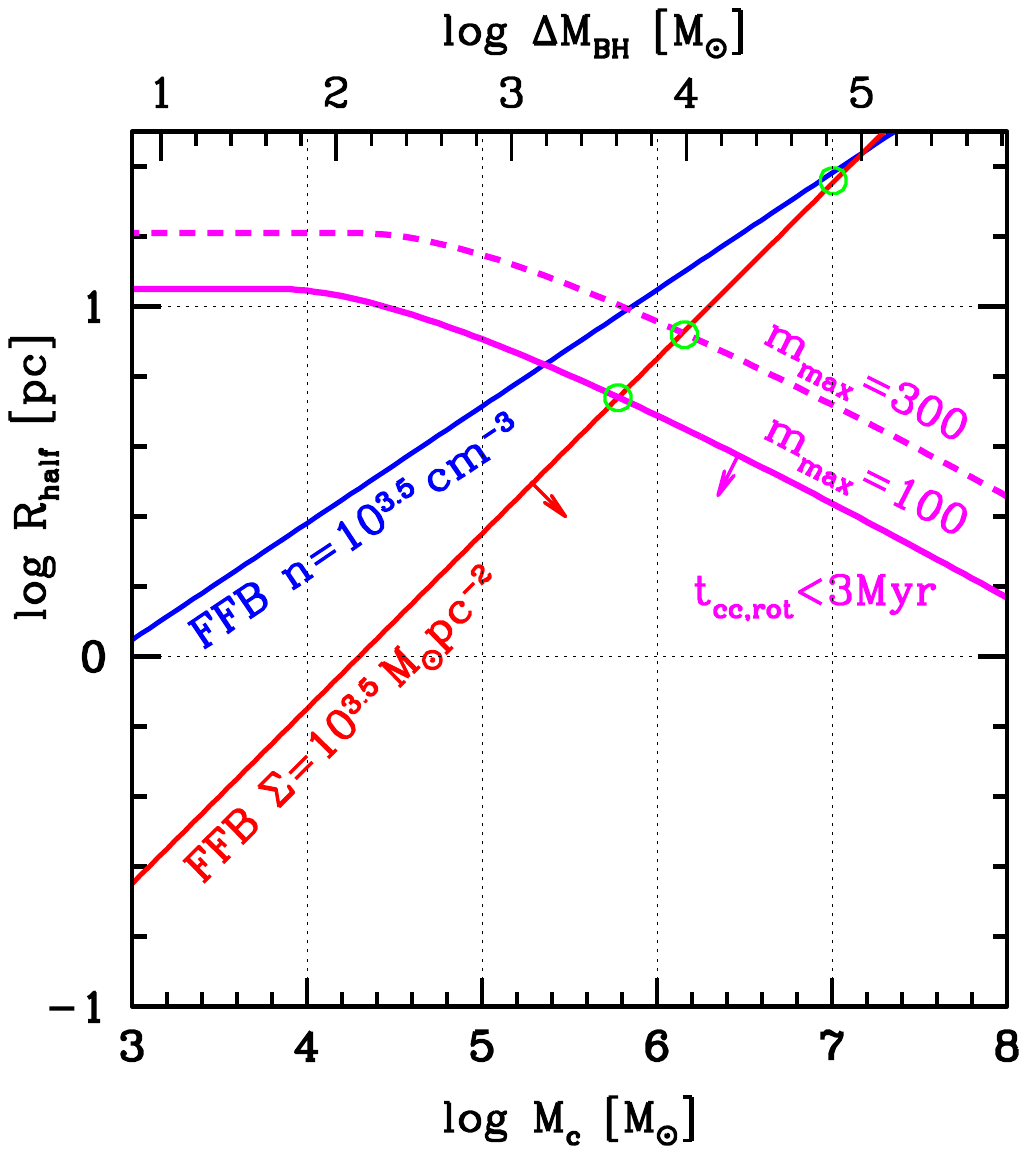}
\vspace{-15pt}
\caption{
Black hole growth in rotating FFB clusters with a broad IMF.
Same as \fig{bh_imf}, right panel, based on \equ{tcc_rizzuto}
\citep[][eq.~4]{rizzuto21},
but with $\tcc$ reduced by a factor of 10 due to rotation
\citep[gravo-gyro instability;][]{hachisu79,hachisu82,hong13,kamlah22}
and flattening.
The FFB clouds that form by disk instability at the characteristic
Toomre mass of $\sim 10^7\msun$ \citep{dekel23},
lie above the curve (magenta) that marks core-collapse in $3\Myr$.
However,
if sub-Toomre clouds form by fragmentation with the same $\Sigma$, they
should lie along the red line, such that core collapse will occur in $3\Myr$
for $\Mc \leq 10^6\msun$.
The corresponding black-hole mass at the end of the FFB phase, \equ{mbh_port},
is $\Mbh \leq 10^4\msun$ (top axis).
}
\vspace{-10pt}
\label{fig:bh_rot}
\end{figure}

\subsubsection{Flattening and dynamical friction}

The simulations above assume little or moderate flattening for the initial 
rotating clusters, of axial ratio up to 0.4, comparable to the typical
flattening of Toomre clumps in simulations of disks at cosmic noon
\citep{ceverino12}.
If the clusters are flatter, with $h/r\ssim 0.2$, say, similar to the parent
galaxy, this would speed up the core-collapse via the shorter
timescale for two-body relaxation and dynamical friction of massive stars.
Based on our estimate below for the dynamical friction 
in a hot disk, \equ{adf_hot}, the effect of flattening could be 
$\tdf \sprop (h/r)^2$, namely an order of magnitude.

Combining the effects of rotation and flattening, we crudely assume
in our feasibility study 
a speed up factor of 10 compared to the non-rotating core collapse time of 
\equ{tcc_rizzuto}. One should be cautioned that this is a tentative crude
estimate, to be explored further by future, more detailed simulations.

\subsection{Core collapse in FFB clusters}

\Fig{bh_rot} shows the threshold in the clump radius-mass 
plane for core collapse in $3\Myr$ in a rotating cluster, to be compared to 
\fig{bh_imf} for a non-rotating cluster. 
The curves, shown for two different IMFs, are
computed using the segregation time in \equ{tcc_rizzuto},
the one used in the right panel of \fig{bh_imf}, but here 
divided by 10 due to clump rotation and flattening.

The maximum clump mass in a disk is assumed to be the characteristic 
Toomre mass as estimated in \citet{dsc09}.
In the FFB scenario, it is derived from the maximum gas mass in the disk 
during each FFB generation of $\sim\!10^9\msun$ to be \citep[][eq.~58]{dekel23}
\be
\MT \simeq 9.2\times 10^6 \msun\, \lambda_{.025}^{5/2}\, \Mveight^{1.14}\,
(1+z)_{10} \, ,
\label{eq:Mt}
\ee
where $\lambda \seq 0.025\,\lambda_{0.025}$ 
is the inverse contraction factor from halo to disk 
(comparable to the halo spin parameter)
and $\Mv \seq 10^{10.8}\msun\,\Mveight$ is the galactic halo virial mass at 
$1\splus z \seq 10\,(1+z)_{10}$.
Below this mass, we assume a distribution of clump masses following
a power-law mass function 
\be
\frac{dN}{dm} \prop m^{-\alpha} \, ,
\label{eq:phi}
\ee
with $\alpha \ssim 1.8$,
as seen in simulations at cosmic noon \citep{mandelker14,mandelker17}.
Such a mass function with $\alpha \ssimeq 2$ is a generic result in a
supersonic turbulent medium \citep{hopkins13,trujillo19,gronke22}.
The fraction of mass in clumps less massive than $\Mc$ is
$f(<\Mc) \seq (\Mc/\MT)^{2-\alpha}$.

Assuming that the clumps of all masses are at the critical surface density for 
FFB, $\Sigma \seq 3\stimes 10^3\Msun\pc^{-2}$ \citep{dekel23,menon23},
one can see in \fig{bh_rot} that the most massive clumps of $\sim\!10^7\msun$
are above the threshold line, namely their core collapse takes longer than
than $3\Myr$.
However, rotating clumps of $\sim 10^6\msun$ and less
should core-collapse in less than $3\Myr$, namely during their FFB phase.
Given the assumed clump mass function, 63\% of the mass is in clumps below
$10^6\msun$, and 23\% is in clumps between $10^5$ and $10^6\msun$.
Based on \equ{mbh_port},
the BHs in FFB clusters of $\sim\!10^6\msun$ should be $\sim\!10^4\msun$,
which we adopt as our fiducial values.
Recall that certain subsequent growth is expected for BHs below this mass due 
to stellar capture, following \equ{MbhMc}, so the seed mass of
$\sim\!10^4\msun$ should be considered a conservative estimate. 
We thus consider for the stages of evolution following the FFB phase 
a compact disk galaxy consisting of thousands of star clusters, 
with a characteristic mass $\sim\! 10^6\msun$, 
each containing a central intermediate-mass BH seed of $\sim\! 10^4\msun$. 

\begin{figure*} 
\centering
\includegraphics[width=0.99\textwidth] 
{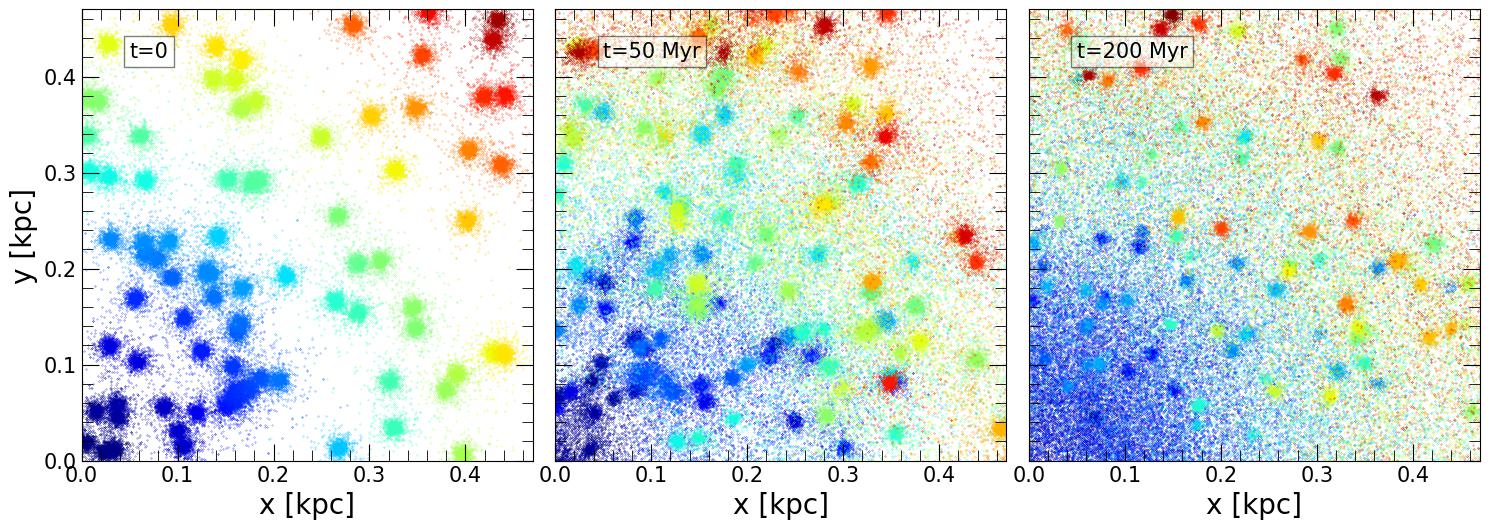}
\caption{
N-body simulation of post-FFB cluster disruption and the buildup of a smooth 
stellar disk component capable of exerting dynamical friction on the BHs 
of each cluster.
Shown is the projected stellar distribution in one quadrant of the galactic 
plane, in a slice of thickness $\pm\!50\pc$, at different times.
One thousand star particles are selected at random from each cluster.
The colors represent radial distance at $t\seq 0$.
The clusters survive mostly intact for a few tens of Myr, namely several 
cluster free-fall times, allowing FFB and core collapse to seed BHs in them.
A significant smooth component is building inside-out to become dominant 
by $\sim\!100\Myr$, after a few galactic orbital times.
}
\label{fig:particles}
\end{figure*}

\begin{figure} 
\centering
\includegraphics[width=0.49\textwidth] 
{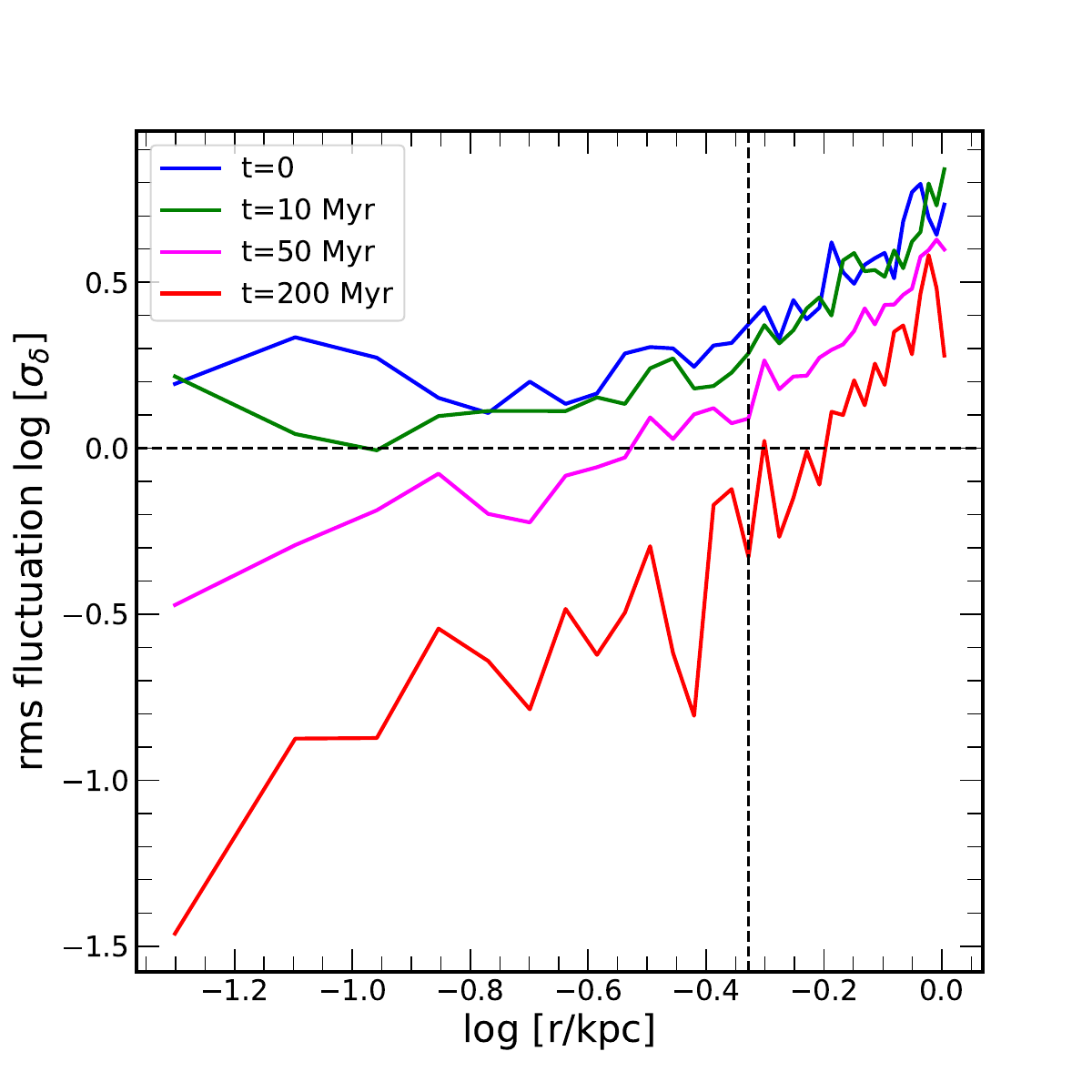}
\vspace{-15pt}
\caption{
The buildup of the smooth stellar disk component.
The presence of clusters is measured by the standard deviation of the density
fluctuation $\sigma_\delta$ in the disk plane near radius $r$ at time $t$.
The density is measured in pixels of side $30\pc$, comparable to the 
original cluster size.
The smooth component becomes significant, with $\sigma_\delta \slt 1$,
after $\sim\!10\Myr$, a few FFB cluster free-fall times, 
and before $\sim\!100\Myr$, several galactic orbital times.
}
\label{fig:sigma_rt}
\end{figure}

\section{Migration of seed BHs by Dynamical Friction}
\label{sec:DF}

The BH seeds, that have been produced in the FFB clusters at cosmic dawn
as described above,
are subject to dynamical friction by the compact galactic stellar system
and the dark-matter halo, which makes them migrate to the galactic center
where they can coalesce into a SMBH.
The question one wishes to verify is whether 
the SMBHs can grow this way by $z \ssim 4\sdash 7$ to $\sim\!10^{7\pm 1}\msun$
keeping a high $\fbh$ of $\sim\!0.01$, as observed.
This requires rather efficient inward migration such that most of the seed 
BHs reach the SMBH by that time, namely in less than a Gigayear.
\adr{
We address this DF-driven inward migration here.
}

\subsection{Disruption of Clusters into a Smooth Medium}

What is the origin of the particles that exert most of the dynamical
friction on the BHs?
The compact configuration of thousands of FFB star clusters is expected 
to turn rather quickly into a smooth galactic stellar system.
This is unavoidable due to efficient tidal disruption by the mutual 
unbound cluster encounters, where clusters of internal velocities of 
$\sim\!10\kms$ orbit with higher relative velocities of $\sim\!50\kms$ in 
a compact thick disk configuration.

We study the evolution of such a system via an N-body simulation using the 
GADGET-4 code \citep{springel21}.
The initial configuration of this idealized experiment consists of $2,000$ 
star clusters of $10^6\msun$ each, randomly distributed in a galactic thick 
disk following an exponential surface-density profile.
To match a typical FFB galaxy at $1\!+\!z\seq 10$ with an integrated SFE
$\epsilon \seq 0.2$ \citep{dekel23,li24},
the total stellar mass is $\Md \seq 2 \stimes 10^9\msun$ 
and the disk half-mass radius is $\Re \seq 470\pc$.\footnote{This radius is 
actually on the high side of the FFB-predicted size distribution, making
the simulated timescale for cluster disruption a conservative overestimate.}
The density profile perpendicular to the disk is exponential
with a scale height of $60\pc$ above and below the plane.
The disk is embedded in an analytic spherical dark-matter halo
of virial mass $\Mv \seq =10^{10.8} \Msun$ and virial radius 
$\Rv \seq 12.3\kpc$, as predicted for typical FFB galaxies at $1\!+\!z\seq 10$. 
The halo density profile is NFW \citep{nfw97} with a concentration parameter
$C\seq 10$.
The equilibrium initial conditions for the N-body disk are generated using 
the AGAMA code \citep{vasiliev19}, where the clusters are tentatively
represented by 2000 point particles.
The cluster initial velocities are drawn from a quasi-isothermal distribution.
The clusters are assigned circular velocities that grow from $\Vc \ssim 50\kms$ 
at $r \seq 0.1\kpc$ to $\Vc \ssim 185\kms$ at $r \seq 1 \kpc$ and beyond.
The three components of the velocity dispersion decline from 
$\sigma_r \ssim \sigma_\phi \ssim 45 \kms$ and $\sigma_z \ssim 65\kms$
at $r \seq 0.1\kpc$ to $\sigma_x \ssim 15 \sdash 20\kms$ at $r \seq 1 \kpc$ 
and beyond.
In a test run with the clusters represented by point particles, 
the density profiles remained stable for at least a Gyr, 
and so did the profiles of mean velocities and $\sigma_z$,
while the profiles of $\sigma_r$ and $\sigma_\phi$
increase after about $0.5\Gyr$ to the level of $30\kms$ beyond 
$r \ssim 0.5 \kpc$.

Each star cluster of $10^{6}\msun$ is assigned a half-mass radius of $7\pc$, 
as predicted for the typical clusters in FFB galaxies \citep{dekel23},
and is modeled as a Plummer sphere in equilibrium 
\citep{plummer11} with 10,000 particles.
The Plummer-equivalent gravitational softening in the simulation is $0.7\pc$. 
This softening was selected based on an equilibrium test, in which
a single cluster of $10^4$ particles was found to be stable over $200\Myr$
with a softening of order 10\% of the cluster half-mass radius.

The simulation of 2,000 clusters with $10^4$ particles in each
is run for $200\Myr$, where the disk dynamical time, 
$\td \seq r/\Vc$, is $2.96\Myr$ at $\Re$ and $4.92\Myr$ at $2\, \Re$.

\Fig{particles} shows a face-on projection of a subsample of the stellar 
particles at three different times in the simulation.
One can see that 
the initial distribution of clusters gradually evolves inside-out into a smooth
stellar disk, capable of exerting dynamical friction on the BHs that
formed at the center of each cluster.  
By $t\seq 50\Myr$, the inner galaxy seems to contain a significant
smooth component. 
By $t \seq 200\Myr$, the whole disk seems to be
dominated by the smooth component.


\Fig{sigma_rt} illustrates the inside-out transition of the FFB galactic disk 
from an assembly of thousands of compact star clusters with central BHs
into a smooth thick stellar disk capable of exerting dynamical friction
on the BHs that formed within the clusters and are left naked
after the cluster disruption.
The fluctuative nature of the density is measured here by the standard 
deviation of the density fluctuation in the disk plane near radius $r$ 
at time $t$. 
The density is measured in pixels of side $30\pc$, comparable to the original 
cluster size, in order to pick up the signal from the clusters while they are
intact.  

We first learn that most clusters survive intact for at least $\sim\!10\Myr$,
namely for a few cluster dynamical times and a typical FFB single generation 
period, 
thus allowing for FFB within each cluster and the core collapse to a seed BH. 
As long as the clusters remain intact, they are subject to dynamical friction.  
The dynamical-friction timescale for a cluster of $10^6\msun$ due to the
dark-matter halo is estimated to be $\sim\!10\Myr$ at $r \seq 30\pc$
and $\sim\!100\Myr$ at $r \seq 70\pc$.
This indicates that a fraction of the clusters that originated from the central
regions of the disk would migrate to the center before they are disrupted.

Second,
we see that the rms fluctuation becomes smaller than unity after about three
orbital times at the half-mass radius,
and smaller than 0.5 over most of the disk mass after about ten
orbital times.
This indicates the presence of a significant smooth stellar component
capable of exerting dynamical friction on the BHs
soon after the FFB phase of the galaxy.
For a typical FFB phase that began at $\zffb \ssim 9$ ($t \ssim 540\Myr$) 
and lasted for $\sim\!100\Myr$, we thus expect the smooth component to be 
dominant for at least $\sim\!500\Myr$ prior to $z \ssim 5$ 
($t \ssim 1150\Myr$).
We find below that the dynamical friction exerted on the BHs by this stellar 
disk is stronger than the dynamical friction exerted by the dark-matter 
halo, and evaluate the corresponding timescale for migration to the galactic 
center.
The dynamical friction by the halo is not included in the above simulation.

We comment that our estimate of the timescale for the formation of a smooth 
disk is likely an overestimate. This is because the clusters, as they undergo 
core collapse associated with mass segregation, are also expected to undergo 
significant evaporation due to two-body effects within the clusters. 
This evaporation is suppressed in our simulations, as we tuned the softening 
length such that each cluster is stable when simulated in isolation.

\subsection{Dynamical Friction in a 3D medium}

While dynamical friction is understood as a complex non-local process involving
resonances \citep{tremaine84,kaur18,banik21}, 
a useful qualitative approximation can be obtained via the classical more local
\citet{chandrasekhar43} formalism, which sums up the two-body
interactions between the moving massive object and the much less massive 
particles in the 3D medium.
Consider a seed BH of mass $m$ on a circular orbit with velocity $V(r)$ 
at a radius $r$ in a disk within an FFB galaxy (disk and halo) whose properties
were determined at $\zffb$ \citep{dekel23,li24}.
The DF acceleration by a spherical mass distribution 
of low-mass particles with a density profile $\rho(r)$ 
and an isotropic Maxwellian distribution of velocities
is approximated by 
\be
\adf(r) = - 4\pi G^2\, {\cal B}(x)\, \ln\Lambda(r)\, 
\frac{\rho(r)\, m}{V(r)^2}\, ,
\label{eq:chandra1}
\ee
where
\be
{\cal B}(x) = {\rm erf}(x)- \frac{2x}{\sqrt{\pi}} e^{-x^2}\, ,
\quad
x = \frac{V}{\sqrt{2}\sigma} \, ,
\ee
and where in the Coulomb logarithm 
\be
\Lambda \ssim \frac{b_{\rm max}}{b_{\rm min}} \ssim \frac{M(r)}{m}\, ,
\ee
is derived by integrating over the distribution of orbit impact parameters.
The estimate of $M(r)/m$ arises from assuming for the maximum and minimum
impact parameters $b_{\rm max} \ssim r$ and 
$b_{\rm min}\ssim Gm/V^2$, and using $V^2 \seq GM(r)/r$.
The DF acceleration then becomes
\be
\adf = - 4\pi\,F\,G\,\rho\,r\, \frac{m}{M(r)} \, ,
\label{eq:chandra2}
\ee
where $F \sequiv {\cal B}\,\ln\Lambda$.

The timescale for migration into the center can be 
obtained by integrating the angular-momentum loss due to the torque
$dL/dt = \tau = -m\,r\,\adf$.
It can be approximated, as demonstrated by following the
orbit from $r$ to the center in an isothermal sphere \citep[][eq.~8.13]{bt08}, 
by
\be
\tdf \simeq \frac{1}{2}\frac{L}{dL/dt} = \frac{1}{2}\frac{V}{\adf}  
= \frac{1}{2F} \frac{M(r)}{m}\, \frac{V/r}{4\pi G \rho} 
= \frac{1}{2F} \frac{M(r)}{m}\, \td \, .
\label{eq:tdf1}
\ee
Here $M(r)$ is the total mass within $r$, and the circular velocity
is approximated by $V^2 \seq G M(r)/r$ even if there is a contribution of 
a disk to the mass and to the circular velocity 
(this can be made more accurate).
The last term is the dynamical crossing time of the system 
(as $V/r\seq \td^{-1}$ and $(4\pi G\rho)^{-1} \seq \td^2$).

\subsection{Dynamical Friction in a Disk}
\label{sec:DF_disk}

An analogous derivation of dynamical friction in a disk is apparently less
straightforward and, quite surprisingly, there is no accepted textbook
approximation analogous to the Chandrasekhar formula, neither based on analytic
integration nor using simulations
\citep[e.g.][]{quinn86,donner93,bekki09}.
We appeal here to our own crude analytic estimates, and test their qualitative
validity with simple N-body simulations.

\subsubsection{Analytic estimates of Dynamical Friction}

Consider an object of mass $m$ moving in a circular orbit at radius $r$
within a disk that consists of much less massive particles, 
with a surface density profile $\Sigma(r)$ and half-thickness $h(r)$. 
The circular velocity $\Vc(r)$, and the associated angular-velocity
$\Omega(r)$, are determined by the mass distribution in the
disk and possibly an additional component, such as a bulge or a DM halo,
whose direct contribution to the dynamical friction is neglected.
We crudely approximate $\Vc^2 \seq G\,M(r)/r$,
where $M(r)$ is the total mass within a sphere of radius $r$,
and tentatively assume for simplicity a flat rotation curve 
$\Vc(r) \seq {\rm const}$.
We use the standard result that 
the disk is self-regulated to comparable kinematic and spatial axial ratios,
$V/\sigma \seq r/h$, where $\sigma$ is the radial
velocity dispersion. This can be derived from hydrostatic equilibrium 
in the vertical direction assuming an isotropic velocity dispersion.

Building upon \equ{chandra1} and \equ{chandra2},
we adopt the Chandrasekhar expression for the magnitude of the DF acceleration,
\be
\adf = 4\pi\, \Fd\,\frac{G^2\,\rho\,m}{\Vrel^2} 
= 4\pi\, \Fd\,G\,r\,\frac{\Sigma}{h}\, \frac{m}{M(r)}\,
\left( \frac{\Vc}{\Vrel} \right)^2 \, ,
\label{eq:chandra_disk}
\ee
where $\Vrel$ is now the relevant relative velocity between the object and the 
background particles exerting the force, and $\Fd$ is a proper numerical 
coefficient that results from the integration over the impact parameters 
and replaces the ${\cal B}\,\ln\Lambda$ factor of \equ{chandra1}.

In the case of a kinematically hot, thick disk, or when $m \sll \Md$,
the relative velocity is typically 
\be
\Vrel \simeq \sigma \simeq \Vc \frac{h}{r} \, .
\label{eq:Vrel}
\ee
A similar expression is obtained alternatively from the differential rotation 
as measured in the frame that is rotating with the object.
We realize that the DF is dominated by impact parameters of order $b \ssim h$.
This can be understood as follows.
On the one hand, 
the gravitational force from each particle scales with $b^{-2}$. 
Furthermore, the contribution of differential rotation to DF also scales 
with $b^{-2}$ because the relative velocity is $\Vrel \sprop b$ 
to first order in $b/r$ (or at any $b$ for a flat rotation curve). 
On the other hand, the number of particles contributing to the DF 
increases as $b^2$ when $b \slt h$ due to the 3D nature of the distribution
within the disk, 
while it only increases as $b$ for $b \sgt h$ due to the 2D nature of the 
distribution on scales larger than the disk thickness.  
The key quantity $\Vrel$ in \equ{Vrel} is the differential rotation between 
radii $r \spm h$ and $r$ once $h \sll r$ or when the rotation curve is flat.
The relative velocity in \equ{Vrel} introduces an enhancement factor of 
$(r/h)^2$ in $\adf$.

However, due to the differential rotation,
the rings outside and inside the object instantaneous radius $r$ 
exert dynamical friction of opposite signs,
where particles in the outer (inner) ring move with an angular velocity smaller
(larger) than that of the object, thus exerting deceleration (acceleration).
In a disk of uniform density these contributions would have balanced each other
such that the net dynamical friction vanishes.
However, for a declining density profile, if we write to first order in $h/r$
\be
\Sigma(r \spm h) = \Sigma(r) \pm \Sigma'(r)\, h \, ,
\quad \Sigma' \equiv \frac{d\Sigma}{dr} \, .
\label{eq:Sigma}
\ee
the second term makes a non-vanishing contribution to the dynamical friction.
Substituting \equ{Vrel} and \equ{Sigma} in \equ{chandra_disk}, summing over
the outer and inner rings, we obtain for the net DF deceleration in a hot disk
\be
a_{\rm df,hot} = 8\pi\, \Fd\, G\, \Sigma'(r)\, r\,\left(\frac{m}{M(r)}\right)\,
\left(\frac{r}{h}\right)^2 \, .
\label{eq:adf_hot}
\ee 
We note that for an exponential disk with an exponential scale radius 
$\rexp$ one has 
$\Sigma'(r)\,r \seq -\Sigma\,r/\rexp$, 
which is on the order of the surface density $\Sigma$ in the main body of 
the disk.

The factor $\Fd$ is rather uncertain.
The range of impact parameters can be assumed to be bounded from below by the 
Hill radius, where the outward tidal force is balanced by the object's self 
gravity,
\be
\Rhill = r\, \left( \frac{m}{\gamma\,M(r)} \right)^{1/3} \, ,
\label{eq:Hill}
\ee
where $\gamma$ is of order unity.
It is $\gamma \seq 3$ if $M(r)$ represents a point mass 
but it is smaller for a gradually declining density profile 
and it can vanish for a flat core \citep{dekel03}.
For a very low-mass object, $m \sll M(r)$, one has $\Rhill \sll r$ and 
$\Rhill \slt h$.
For example, with $m \ssim 10^{-5} M(r)$ and $h/r \ssim 0.2$,
one has $h/\Rhill \ssim 10$. 
Assuming $b_{\rm max} \ssim h$,
this range would correspond to $\ln\Lambda \ssim 2.3$, and therefore to $\Fd$
of order unity, but its actual value has to be calibrated by simulations.

In the alternative case of a kinematically cold disk, or for $m$ that is not 
extremely smaller than $M(r)$, one has $h \slt \Rhill$. In this case the 
effective impact parameter is $b \ssim \Rhill$ rather than $b \ssim h$. 
Replacing \equ{Vrel} with $\Vrel \ssimeq \Vc \Rhill/r$ for the contribution of
differential rotation to $\Vrel$, and replacing $h$ with $\Rhill$ in
\equ{Sigma}, we obtain by inserting in \equ{chandra_disk}
that the net DF in a cold disk is
\be
a_{\rm df,cold} = 8\pi\, \Fd\, G\, \Sigma'(r)\, r\,
\left(\frac{m}{M(r)}\right)^{2/3}\, \left(\frac{r}{h}\right) \, .
\label{eq:adf_cold}
\ee
The numerical coefficient $\Fd$ here can be different from the coefficient in
the hot-disk case, and it should be calibrated using simulations.

For a razor-thin disk, \citet{valtonen90} analyzed the dynamical friction
in a Mestel disk, $\Sigma \sprop r^{-1}$ with a flat rotation curve, 
and obtained (their Eq.~5) 
\be
a_{\rm df,thin} = \frac{4\pi}{3 \sqrt{3}}\, G\, \Sigma'(r)\, r\,
\left( \frac{m}{M(r)} \right)^{1/3} \, .
\label{eq:val}
\ee
This is similar to the analytic result by \citet{quinn86} in their Eq.~III.20,
and to their simulations,
but with a coefficient larger by a factor of a few due to a different
analysis of the impact parameters and the inclusion of self gravity in the
simulations.
\citet{valtonen90} found \equ{val} to be in general agreement with their 2D 
N-body simulations, although there is certain enhancement in the simulations
by self-gravity, namely by resonances.
Using \equ{val} to calibrate our \equ{adf_cold}, 
for $m \seq 0.04 M(r)$ corresponding to the $m \seq 0.04 \Md$ 
simulated by \citet{valtonen90}, 
would give $\Fd\,r/h \ssim 1$ near these values of $m$ and at large radii. 
Interestingly, \citet{valtonen90} find (in their Fig.~2a) that with a 
self-regulated velocity dispersion at the level of Toomre $Q \seq 1.4$, 
the dynamical friction is not very different than in the razor-thin disk case, 
being weaker by only $\sim 30\%$.
This is consistent with our finding that \equ{adf_hot} and \equ{adf_cold}
provide estimates in the same ball park for hot and cold disks and relatively
large $m$.

\subsubsection{Testing Dynamical Friction with simulations}

\begin{figure*} 
\centering
\includegraphics[width=0.89\textwidth] 
{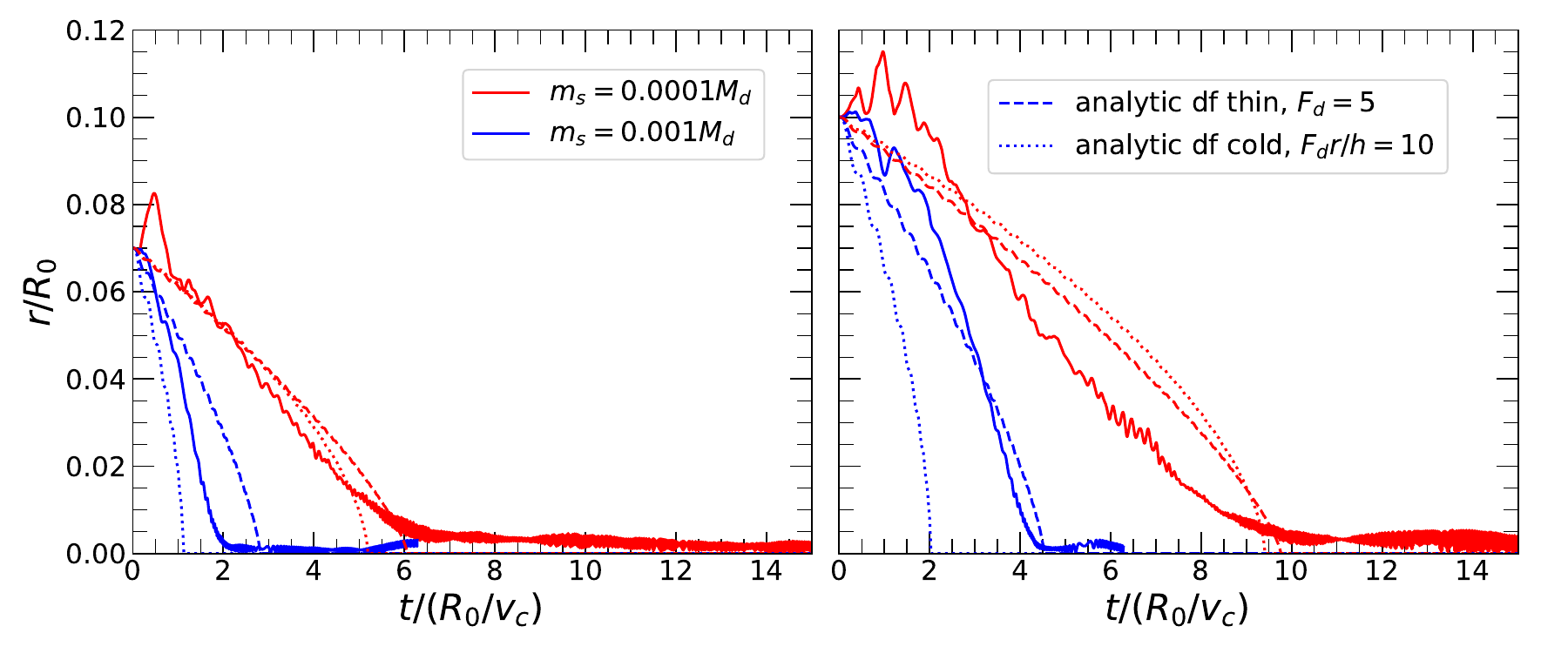}
\vspace{-5pt}
\caption{
Testing the inward migration by dynamical friction in a disk.
The solid curves refer to
N-body simulations of a BH of mass $m$ starting in a circular orbit at $r_i$
within a thin live Mestel disk with an isothermal halo as an external potential.
The left and right panels refer to
$r_i \seq 0.07 R_0$ and $0.1 R_0$.
Shown for comparison are analytic estimates based on \equ{adf_cold} (dotted)
and \equ{val} (dashed), crudely normalized for best overall fit in the four
cases.  We learn from this comparison that the analytic approximations are
realistic conservative estimates.
}
\label{fig:DF_sim}
\end{figure*}

In order to verify the order-of-magnitude validity of the above analytic 
estimates of the dynamical-friction timescale in a disk, we perform our own 
simplified N-body simulations using the GADGET-4 code \citep{springel21}.
We simulate the orbital decay of massive point particles in a razor-thin, cold,
truncated \citet{mestel63} disk, with a surface density profile  
\be
\Sigma(r) = \frac{\Md}{2 \pi R_0} \frac{1}{r}\, 
{\rm cos}^{-1} \left(\frac{r}{R_0}\right) \, ,
\ee
where $\Md$ is the total disk mass and $R_0$ is the disk truncation radius. 
The truncated Mestel disk has a flat circular velocity profile 
$V_{\rm c, disk}^2 \seq (\pi/2)\, G\, \Md/R_0$. 
The disk consists of $10^6$ particles on circular orbits, 
assuming a Plummer-equivalent gravitational softening of $0.001 R_0$.
To help long-term stability, we add an inert isothermal-sphere halo with a 
circular velocity $3\,V_{\rm c, disk}$, such that the total circular velocity
of the system is $\Vc \simeq 3.16 V_{\rm c, disk}$, but this halo does not 
exert additional dynamical friction.

We first verify the stability of the simulated disk for several orbital times,
when it maintains its density profile while developing a modest velocity 
dispersion in the radial and tangential directions at the level of $0.1\Vc$. 
We then simulate the orbital decay of massive point-like objects of two 
different masses, $m \seq 10^{-4} \Md$ and $m \seq 10^{-3} \Md$,
with the same softening as that of the disk particles, 
starting on a circular orbit at radius $r_i$. 
\Fig{DF_sim} shows the radial migration of these massive particles,
starting at either $r_i \seq 0.07 R_0$ or $0.1 R_0$.
The migration timescale in these simulations is found to be shorter by a 
factor $\sim\! 3$ than the time derived by the crude approximation
$\tdf \seq 0.5 \Vc/\adf$ from \equ{val}.
The simulated DF timescale is comparable to the time derived from 
\equ{adf_cold} with $\Fd\,r/h \ssimeq 6$.  
The analytic curves that are shown in comparison to the simulations
in \fig{DF_sim} are obtained by solving for a particle 
orbit under the smooth force due to the enclosed mass at a given radius 
and the expression for the force of dynamical friction.
For a razor-thin disk, the acceleration in \equ{val} is multiplied by a 
numerical constant $\Fd$, which is tuned for a crude simultaneous match to 
the four simulation results. 
For a cold disk, the value of $\Fd r/h$ is tuned in \equ{adf_cold}. 
A reasonable match to all four cases is obtained for a factor $\Fd\seq 5$
multiplying \equ{val},
and for $\Fd r/h \seq 10$ in \equ{adf_cold}, the curves shown in the figure.
The fact that these factors are somewhat different from the fits to $\tdf$
quoted above partly reflects the inaccuracy of the approximation 
$\tdf \ssim 0.5 \Vc/\adf$.

This simple test with simulations indicates that the analytic estimates 
of \equ{adf_hot} and \equ{adf_cold}, with $\tdf \ssim 0.5 \Vc/\adf$,
can be used for our purpose of a crude, conservative
estimate of the dynamical-friction timescale in a cold disk.

\subsection{An FFB Galaxy exerting dynamical friction}

According to the FFB model \citep{dekel23}, 
most FFB galaxies occur at redshifts $\zffb \sgeq 8$ 
in halos near and above a threshold mass of 
\be
\Mveight = (1+\zffb)_{10}^{-6.2} \, ,
\label{eq:threshold}
\ee
where $\Mv \sequiv 10^{10.8}\msun\,\Mveight$ and $1+z \sequiv 10\,(1+z)_{10}$.
The corresponding halo virial radius is
\be
\Rv = 12.3\kpc\, \Mveight^{1/3}\, (1+\zffb)_{10}^{-1}\, .
\label{eq:Rv}
\ee
%

We assume that the BHs are initially distributed within an FFB exponential
stellar disk, which turns out to dominate the dynamical friction. 
The disk mass is 
\be
\Md = \epsilon\,\fb\,\Mv \, ,
\label{eq:Md}
\ee
where the star-formation efficiency $\epsilon$ may range from $0.2$ to unity.
The disk half-mass radius is \citep{li24}
\be
\Re = 0.31\kpc\, \lambda_{.025}\, (1+\zffb)_{10}^{-3.07} \, .
\label{eq:Re}
\ee
The $1\sigma$ scatter in size derives from the scatter in the spin parameter,
$\sigma_{\ln \lambda} \simeq 0.5$, namely a multiplicative factor of 1.6.
The disk profile is assumed to be exponential,
\be
\Sigma(r) = \Sigma_0 e^{-x}\, , 
\quad x = \frac{r}{\rexp} \, ,
\label{eq:exp_Sigma}
\ee
with the exponential radius $\rexp \seq \Re/1.68$.
The corresponding disk mass profile is
\be
\Md (r) = \Md [1-e^{-x}(x+1)] \, ,
\quad \Md = 2\pi \rexp^2 \Sigma_0 \, .
\label{eq:exp_mass}
\ee
The gradient, if needed, is
\be
\Sigma'(r) = -\frac{\Sigma(r)}{\rexp} \, .
\ee
Again, we approximate $\Vd^2 = G \Md(r)/r$.

\begin{figure*} 
\centering
\includegraphics[width=0.49\textwidth,trim={1.5cm 5.5cm 1.5cm 4.5cm},clip]
{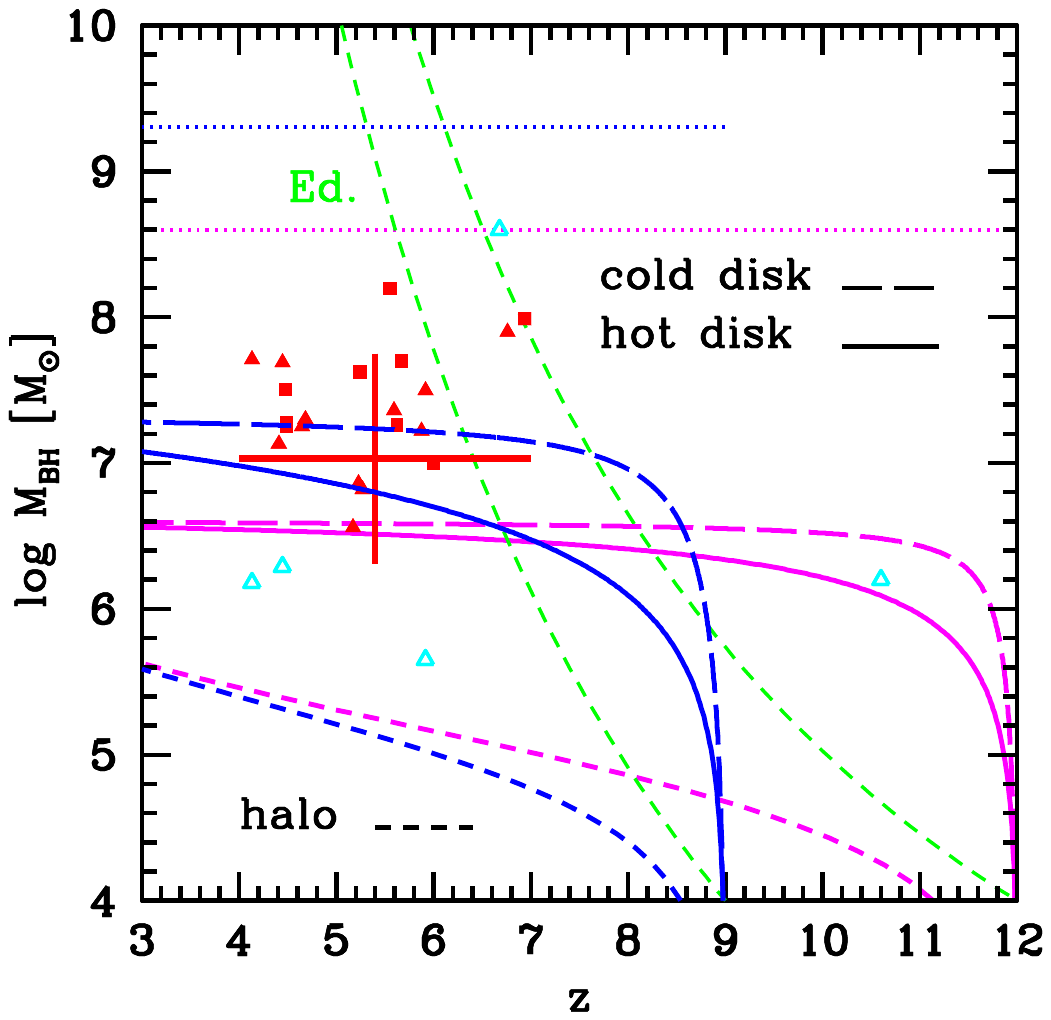}
\includegraphics[width=0.49\textwidth,trim={1.5cm 5.5cm 1.5cm 4.5cm},clip]
{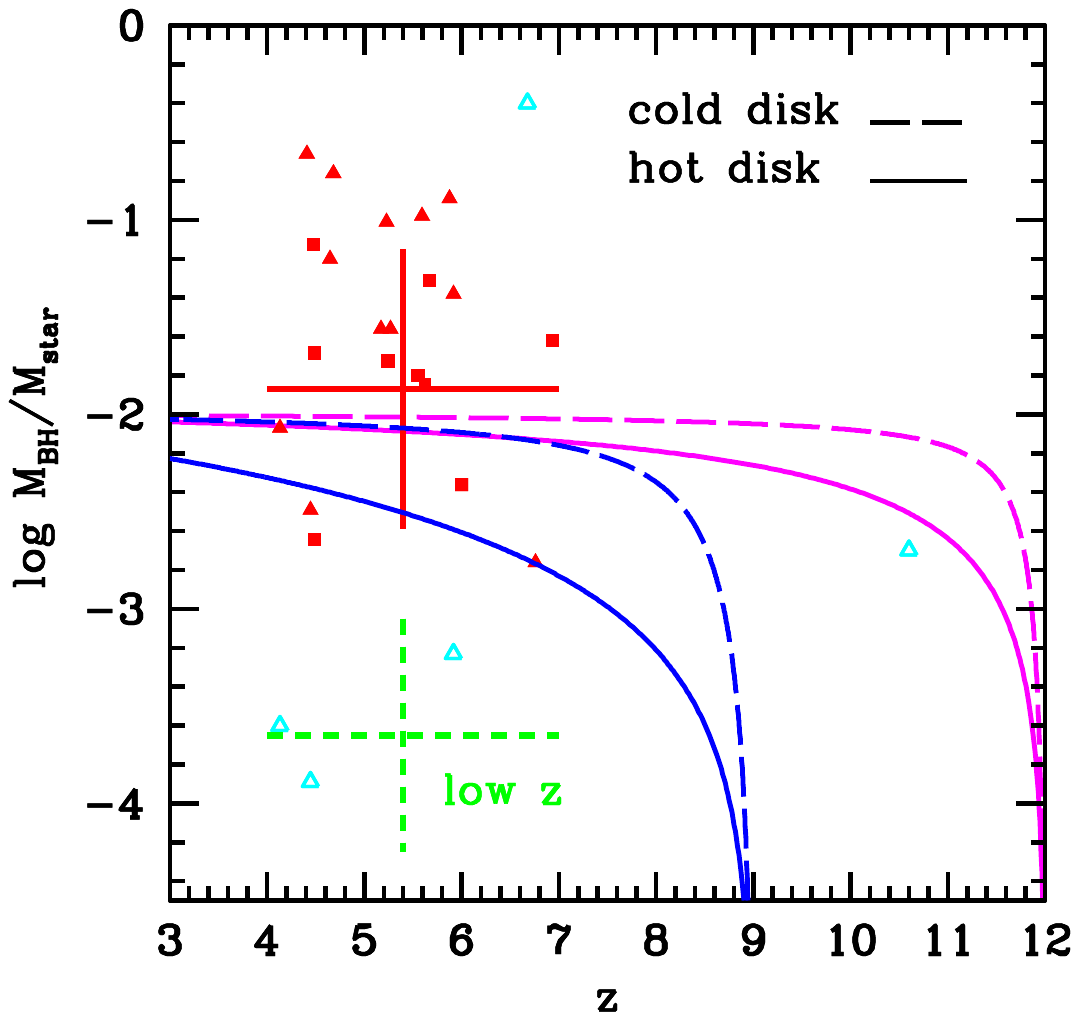}
\vspace{-15pt}
\caption{
SMBH mass (left) and BH-to-Stellar mass ratio (right)
as a function of redshift for $\zffb \seq 9$ (blue) and $12$ (magenta),
with $m\seq 10^4\msun$, either in a hot disk (\equ{adf_hot}, solid)
or in a cold disk (\equ{adf_cold}, long-dashed).
The FFB-threshold halo properties as a function of $\zffb$ are from
\equ{threshold} and \equ{Rv} \citep{dekel23}.  The SFE is $\epsilon \seq 0.2$.
The BHs are assumed to be on circular orbits responding to DF from the stellar
disk ($V/\sigma \seq 5$) plus an NFW DM halo (whose contribution is negligible,
short dashed).
The BHs are assumed to start in an exponential disk whose radius is
estimated in \equ{Re} \citep{li24}.
The galaxy stellar mass, assumed to be constant in time, is marked (dotted).
The $1\sigma$ scatter due to the scatter in disk size in \equ{Re}
is $\pm 0.2$dex.
Eddington growth is marked for comparison (thin dashed green)
The observational estimates are shown for comparison.
The data are from \citet{maiolino23} (circles), \citet{harikane23} and
\citet{ubler23} (squares).
The selection-bias-corrected mean and standard deviation by \citep{pacucci23}
based on the data in red symbols are marked by the red lines.
The values of $\fbh$ at $z\seq 4\sdash 7$ are $>\!3\sigma$ above the
standard $z \seq 0$ low-$z$ ratio by \citet{reines15} (indicated by
dashed green lines).
The average SMBH masses and $\fbh$ ratios as deduced from the observations
at $z\ssim 4\sdash 7$ are reproduced by the $10^4\msun$ seeds of the FFB
scenario.
}
\vspace{-10pt}
\label{fig:smbh}
\end{figure*}

While the contribution of the dark-matter halo to the dynamical friction turns
out to be negligible compared to that of the disk,
as seen in the left panel of \fig{smbh}, 
we include it for completeness.
We assume an NFW dark-matter halo profile
\be
\rho_{\rm h}(r) = \frac{\rho_{\rm s}}{x\, (1+x)^2} \, ,
\ee
where $x \seq r/r_{\rm s}$ and the concentration parameter is
$C \seq \Rv/r_{\rm s}$.
We adopt $C\seq 4$ at high $z$ \citep[][Fig.~20]{zhao09}.
The NFW mass profile is
\be
\Mh(r) = 4\pi\rho_{\rm s}\,r_{\rm s}^3\, A(x) \, ,
\quad
A(x) = \ln(x+1) - \frac{x}{x+1} \, .
\ee
With the free parameters $\Mv$ and $C$, we have
$r_{\rm s} \seq \Rv/C$ and $\rho_{\rm s} \seq \Mv/[4\pi r_{\rm s}^3 A(C)]$.
The halo contribution in quadrature to the circular velocity is
$\Vh^2 \seq G\Mh(r)/r$.

\adr{
We very crudely assume here that the galaxy does not grow from the FFB phase 
to the epoch when the SMBH is detected, an assumption that may cause an
overestimate in $\fbh$. The average halo growth between 
$z \seq 9$ and $z \seq 7$ is expected to be by a factor of 
$\exp [0.8*(9-7)] \ssim 5$ \citep{dekel13}, namely by less than an order of 
magnitude. This can serve as a conservative upper limit for the potential
growth of stellar mass.
The large stellar mass produced already in the FFB phase, and the 
evidence from {\tt JWST} for rapid quenching soon after cosmic dawn  
\citep{degraaff24, weibel24, antwidanso24}, 
support the simplified assumption of a more limited growth of stellar mass. 
This assumption is to be relaxed in future cosmological simulations.
}

\subsection{Inspiraling of BH seeds and Merger Rate into an SMBH}
\label{sec:merger_rate}

We assume that the seed BHs of mass $m$ at $\zffb$ are distributed in the
FFB disk of mass $\Md$, with values of $\fbh$ 
that are determined by the core collapse in the FFB clusters. 
The total BH mass is thus $\fbh\, \Md$.

At any given radius $r$, we compute the density, mass and circular
velocity profile of the disk and the halo and their sums.
The circular velocities are approximated assuming spherical symmetry.
The DF acceleration at $r$ is computed from the sum of the contributions
of the two components,
and the DF time $\tdf(r)$ is derived from it.
The SMBH mass at a Universal time $t \seq t_{\rm ffb} + \tdf(r)$
is assumed to be the sum of the BH seed masses that were initially 
in the disk at radii smaller than $r$ (ignoring any mass loss by recoils,
to be discussed in \se{recoil}).

At a Universal time $t$, we approximate the redshift by 
\be
(1+z) = (t/t_1)^{-2/3} \, ,
\ee
where $t_1 \seq 17.5\Gyr$ \citep{dekel13}. 
This is used at $\zffb$ and at any other $z$.

\begin{figure} 
\centering
\includegraphics[width=0.49\textwidth,trim={1.5cm 5.5cm 1.5cm 4.5cm},clip]
{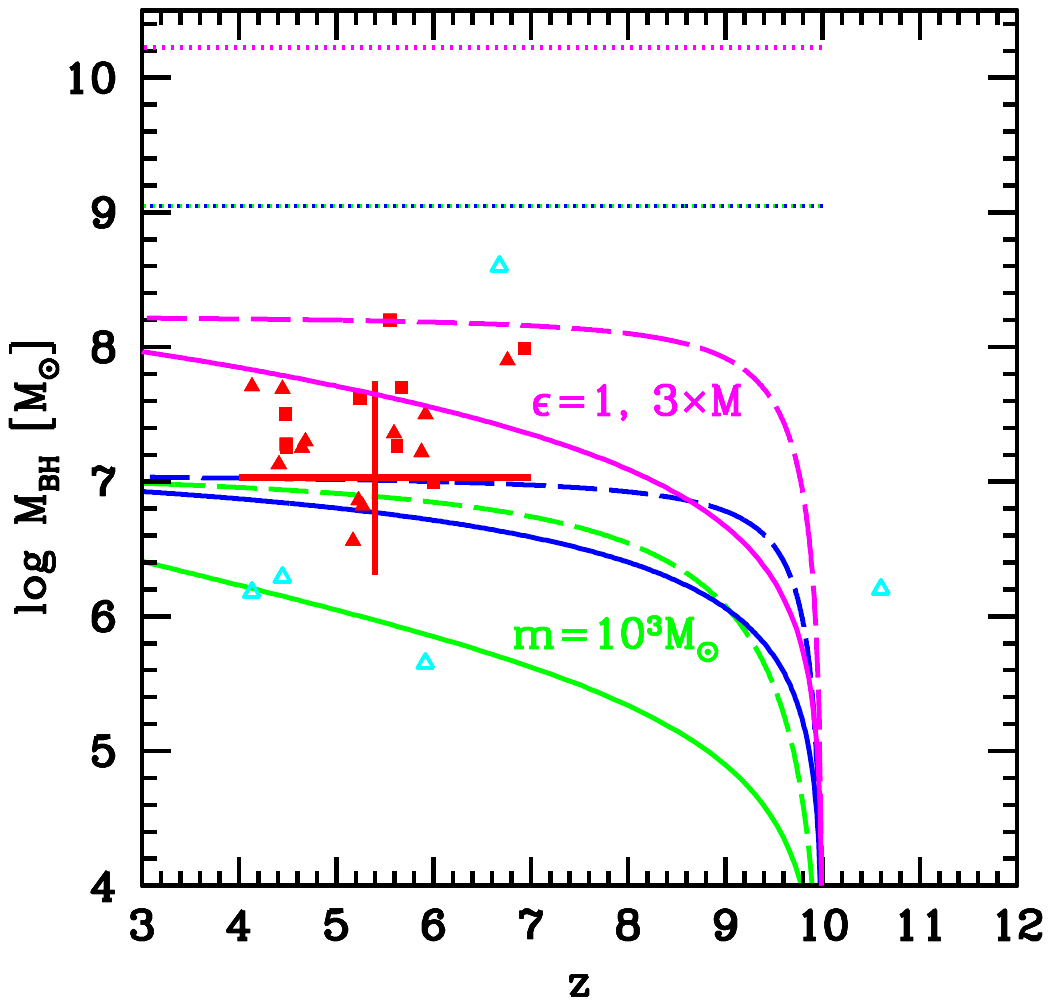}
\vspace{-15pt}
\caption{
Scatter in SMBH mass.
Same as \fig{smbh} but for $\zffb \seq 10$ and for three caes:
{\bf Middle, blue:}
the fiducial case, with $\epsilon \seq 0.2$, halo mass at the FFB threshold,
and seed BHs of $m \seq 10^4\msun$, is recovering the average BH mass at
$z\seq 4\sdash 7$.
{\bf Top, magenta:}
with $\epsilon \seq 1$ and halo mass 3x the FFB threshold,
the model reproduces the most massive BHs at $z\seq 4\sdash 7$.
{\bf Bottom, green:}
with seed BHs of $m\seq 10^3\msun$, the BH growth in a hot disk is not
sufficiently fast.
}
\vspace{-10pt}
\label{fig:smbh_m}
\end{figure}

The predicted SMBH mass at redshift $z$ and the corresponding BH-to-stellar
mass ratio are shown in \fig{smbh} and \fig{smbh_m} as a function of $z$,
in comparison to the observational estimates at $z \seq 4\sdash 7$.
This is under the tentative assumption that the BH seeds that reach the 
central region of the galaxy by DF-driven migration indeed merge efficiently
into a central SMBH (to be reconsidered in the following section).
In \fig{smbh} the seed mass is assumed to be $m\seq 10^4\msun$ 
based on \fig{bh_rot}, with $\fbh \ssim 0.01$ based on \equ{mbh_port}.
Shown are the predictions for FFB at $\zffb \seq 9$ and $12$.
\Fig{smbh_m} illustrates the scatter in the SMBH growth via different cases of 
BH seed mass and galaxy mass above the FFB threshold, for $\zffb \seq 10$.
The dynamical friction in the disk is estimated either by \equ{adf_hot} 
(hot disk, solid curves) or by \equ{adf_cold} (cold disk, dashed curves).
The FFB galaxy stellar mass, which is very crudely assumed here to remain 
constant since $\zffb$, is shown for comparison (dotted line) in the panels 
showing the SMBH mass.
The FFB-threshold halo properties that are adopted in the calculation
as a function of $\zffb$ are from \equ{threshold} and \equ{Rv} \citep{dekel23}.
In the fiducial case, the assumed integrated SFE is $\epsilon \seq 0.2$, 
based on the tentative best fits reported in \citet{li24} between the FFB 
predictions and a variety of JWST observations of galaxies at cosmic dawn. 
This value of average SFE indeed reflects the predicted duty cycle of the 
star-formation history during the $\sim\!100\Myr$ FFB phase 
\citep[][Fig.~9]{li24}.
The BHs are assumed to be on circular orbits responding predominantly 
to the DF from the stellar disk (with $V/\sigma \seq 5$ in \equ{adf_hot} or
\equ{adf_cold}). 
The additional contribution from an NFW halo turns out to be negligible.
The BHs are assumed to start in an exponential disk whose radius is
estimated in \equ{Re} \citep{li24}.
An additional $\pm 1\sigma$ scatter due to scatter in the disk size
in \equ{Re} corresponds to $\pm 0.2$dex.

We learn from \fig{smbh} that
if the DF in the disk is acting according to the estimate
for a hot disk, \equ{adf_hot}, the fiducial model with $m \ssim 10^4\msun$ 
reproduces the median $z\seq 4\sdash 7$ observations both in terms of BH masses 
and the high $\fbh$ ratios.  
We learn further from \fig{smbh_m} that in the case of a cold disk, 
\equ{adf_cold}, even more modest seeds of $m \ssim 10^3\msun$ could make it to 
the SMBH in time, starting from FFB at $\zffb \ssim 10$.

The most massive BHs in the $z\seq 4\sdash 7$ sample, of $\sim\!10^8\msun$,
can be reproduced in post-FFB galaxies at the tail of the distribution of the
model parameters, e.g., if the halo mass is a few times the threshold mass 
for FFB, or if the SFE is at the high end toward unity, or if the optimal 
conditions for SMBH growth are pushed to their limit.  
These include a top-heavy IMF and strong rotational support that speed-up 
the core collapse to seed BHs in the FFB star-forming clumps, 
a cold galaxy disk configuration with a steeply declining surface density 
and angular-velocity profiles for rapid DF-driven inward migration,
and a cold disk of BHs for spin-orbit alignments that minimize the suppression
of BH growth by GW recoils (see the following section).
In addition, as discussed below in \se{compaction}, 
strong wet compaction events (e.g. due to galaxy major mergers) can boost the 
SMBH growth to its extreme efficiency. 
\adr{
We also recall that the merger-driven scenario is only one of the
mechanisms for SMBH growth; it is likely to be accompanied by 
central accretion-driven growth via an accretion disk,
which can boost the BH mass and may possibly 
increase $\fbh$ to higher values. 
}

\adr{
The mass range of galaxies that host SMBHs is expected to be rather broad, as
observed at $z \seq 4 \sdash 7$.  
FFBs are expected in all galaxies above a threshold halo mass that 
decreases steeply with redshift, \equ{threshold},
based on Eqs.~62 and 67 and Fig.~6 of \citet{dekel23}.
Therefore, a range of values for the epoch of FFB, $\zffb$, is expected to lead
to a wide range of stellar masses for galaxies that host over-massive SMBHs, 
ranging from below $10^8\msun$ to above $10^{10}\msun$.
}

\adr{
Over-massive BHs can be produced at higher redshifts, out to $z \ssim 10$ and
beyond, as observed in a few cases \citep{bogdan24, kovacs24, maiolino24}, 
in galaxies that underwent FFB at even higher redshifts.  
This is demonstrated in \fig{smbh}, which also shows the estimated BH mass 
and $\fbh$ for a galaxy that underwent FFB at $\zffb \seq 12$. 
It shows that the high values as observed at $z \ssim 10$ can be obtained even 
for hot galactic disks, in galaxies of a stellar mass of 
$\sim\!4 \stimes 10^8\Msun$.
FFB galaxies that formed at $\zffb \seq 15$ would lead to over-massive BHs 
in galaxies of stellar mass as low as $\sim\! 10^8\Msun$.
}

\adr{
A note of caution is that the DF-driven inward migration of the BHs may be 
suppressed by scattering against the irregular stellar distribution. This is 
indicated in zoom-in cosmological simulations of BHs in high-$z$ dwarf 
galaxies, utilizing a subgrid model for capturing the unresolved dynamical 
friction acting on the BHs \citep{pfister19}.
This emphasizes the need for the post-FFB galaxies to be rather cold disks in 
order to make the sinking into the central regions efficient. It adds to
the requirement of cold disks in order to avoid BH ejections by post-merger 
recoils, to be discussed in \se{recoil}. On the other hand, spiral arms that 
can form in a cold disk may weaken the DF-driven migration. 
Challenging, high-resolution simulations of the DF-driven cluster and BH 
migration in a realistic post-FFB disk galaxy are clearly needed in order 
to investigate the validity of DF-driven migration as well as the BH mergers 
and the effect of recoils.
}

\section{Gravitational-Wave Recoil and the Final Parsec}
\label{sec:recoil}

Our prior calculations of SMBH growth in FFB galaxies  
(\se{merger_rate}, \fig{smbh} and \fig{smbh_m}) 
implicitly assume that every BH seed that makes it to the galactic center will 
merge with the central BH, culminating eventually in merger-driven SMBH 
production. This sequence of runaway seed BH mergers parallels the runaway 
stellar mergers that produced the seed BHs in the first place inside their birth
clusters. However, there are, in principle, multiple steps along the way that 
could derail this ``runaway of runaways'' as the BHs approach the galaxy 
nucleus. One is the `final parsec problem' and the other is the post-merger
recoils, to be examined next.

\subsection{The final parsec problem}
\label{sec:final_pc}

The ``final parsec problem" \citep{begelman80} is thought to
suppress massive BH mergers in galactic nuclei. In brief, a circular-orbit BH 
binary can only merge due to gravitational wave emission in less than a Hubble
time if it is already extremely tight, with a semi-major axis smaller than a
milliparsec.  
However, dynamical friction ceases shrinking the binary orbit once it becomes 
``hard'' with respect to the surrounding stellar population, typically on 
scales $1\sdash 10\pc$. Therefore, if massive BH binaries are to merge in 
galactic nuclei, some additional hardening process must operate on sub-pc 
scales \citep{merritt05}. 

The simplest of these processes is repeated strong three-body scatterings with
stars in the galactic nucleus. Ejection of stars to infinity can in principle 
drain enough energy to shrink the BH binary down to milliparsec scales. 
While this process is relatively inefficient in spherical symmetry 
\citep{milosavljevic03}, it is greatly accelerated in asymmetric potentials 
where stellar orbits fail to conserve angular momentum \citep{merritt04}.
Even a small degree of nuclear triaxiality (only a $\sim\!5\%$ deviation from 
axial symmetry) can efficiently solve the final parsec problem through stellar 
ejections \citep{vasiliev15}. We speculate that this triaxiality should arise 
naturally in the messy and dynamically young environments of FFB galaxies. 

Furthermore,
the BHs migrating to the center are likely to carry with them 
clumps of stars from the original FFB clusters. When such systems merge,
the stars can help tighten the BH binary, either at the center, or while
spiraling in.

In addition, the presence of significant amounts of nuclear gas may help
solving the final parsec problem via circum-binary torques from active galactic 
nucleus accretion disks \citep{armitage02}. This scenario is interesting 
insofar as accretion from gas disks will independently lead to SMBH growth 
(beyond what is provided from hierarchical BH mergers). We caution, though, 
that as a solution to the final parsec problem, gas torques can be self-limited
by Toomre instability \citep{lodato09}, and in some circumstances can even 
lead to binary expansion \citet{lai23}.
The process of wet compaction discussed in \se{compaction} is a natural way to
bring in the necessary gas into the galaxy nucleus.

If none of the above scenarios efficiently harden the orbits of nuclear BH 
binaries, the secular and chaotic dynamics of BH triples may help. The large 
number of inspiraling BHs means that BH binaries will not live in isolation for
long, and if they do not quickly merge due to gas torques or triaxiality-driven
stellar scatterings, they will soon become part of a BH triple system. If the 
BH triple is hierarchical, the inner BH binary will often be driven to merge 
from Kozai-Lidov oscillations \citep{blaes02}, though in some orbital 
configurations this will not occur. Conversely, if the triple becomes 
non-hierarchical, then strong, chaotic scatterings will result, with three 
possible outcomes, each with order unity probability.  First, a pair of BHs 
may merge promptly \citep{hoffman07, ryu18, bonetti18}.
Second, one or more BHs may be ejected from the system. It is generally the 
lightest BH that is ejected, and this ejection may trigger the merger of 
the remaining pair \citep{bonetti18}. Third, the outcome of strong scattering 
may be to create a hierarchical triple that does not evolve significantly over
a Hubble time.  However, the nature of our FFB galactic system makes it 
unlikely that three-body ejections can derail runaway growth of the central 
object.  Low ejection velocities mean that the ejected BH will usually return 
to the galactic center \citep{bonetti18}, and strongly preferential ejection 
of the lightest object means that a heavy central SMBH cannot be lost in this 
way.  Likewise, ``failed triples'' (situations in which three-body dynamics 
does not solve the final parsec problem) will not be the end of the story as
continued infall of new BH seeds means that eventually, fourth or fifth BHs 
will arrive to trigger chaotic orbital evolution and new strong scatterings.

Given the variety of ways in which our systems can solve the final parsec 
problem, one may argue that this is not likely to be the obstacle to SMBH 
growth by mergers. We therefore focus our attention on a 
potentially more serious bottleneck: GW recoil.

\subsection{Recoil Velocity}

When two BHs merge, the merged BH suffers a recoil caused by the anisotropic 
emission of gravitational radiation. When the recoil velocity exceeds the 
escape velocity from the galaxy, it removes the BH from the central 
regions and suppresses the initial growth of the SMBH by mergers. 
The recoil velocity is robustly estimated using non-linear general 
relativistic calculations \citep{campanelli07,schnittman08,lousto10,gerosa16}.
It is a strong function of the mass ratio $q \seq m_2/m_1 \sleq 1$ 
while it is independent of the absolute values of the masses. 
It also depends sensitively on the spins of the BHs, 
the angles between the spins and the merger orbit, 
and for misaligned cases on the phase of the orbit at the merger
(considered to be a random variable). 

\begin{figure} 
\centering
\includegraphics[width=0.49\textwidth] 
{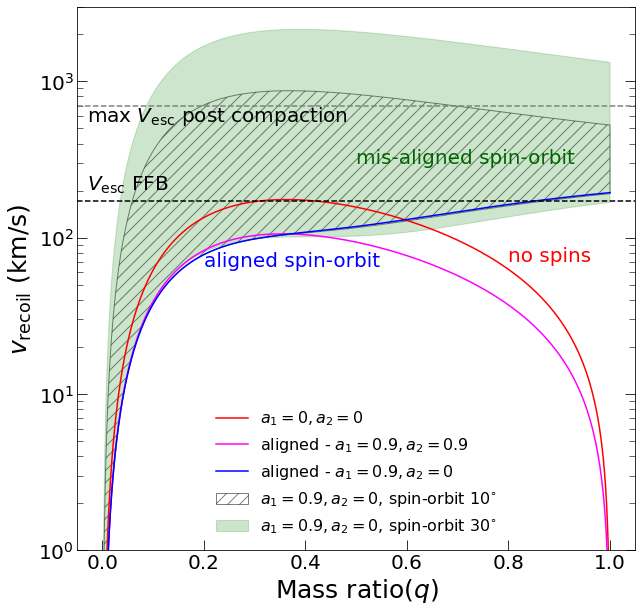}
\vspace{-10pt}
\caption{
Recoil velocity of the merged BH due to a merger between two BHs as a function 
of their mass ratio $q \seq m_2/m_1$.  The cases shown are 
(i) non-spinning BHs (red), 
(ii) a strongly spinning BH $a_1 \seq 0.9$ where the spin is aligned with the 
orbit (blue), 
(iii) strongly spinning BHs $a_1 \seq a_2 \seq 0.9$ (with respect to the maximum
possible $a_1 \seq 1$) with the spins aligned with the orbit (magenta), 
and
(iv) a strongly spinning BH $a_1 \seq 0.9$ with the spin inclined at 
$10^{\circ}$ or at $30^\circ$ with respect to orbit, spanning the full range
of phases for the encounter (diagonal lines area and shaded area,
respectively).
Marked in comparison are the fiducial and maximum escape velocities from 
an FFB galactic disc of $171$ and $700\kms$ (black dashed, gray dashed).
}
\label{fig:spin}
\end{figure}

\Fig{spin} shows the recoil velocity as a function of the mass ratio 
for certain cases of spins.
A fitting formula for non-spinning BHs is provided by \citep{letiec10}
\be 
\Vk = 9.5\, \eta^2 (1-4\eta)^{1/2} (1+ 0.3\,\eta) \times 10^3 \kms \, ,
\label{eq:kick}
\ee
where $\eta$ is the symmetric mass ratio,
$\eta \seq m_1 m_2/(m_1+m_2)^2$.\footnote{In
terms of the mass ratio $q \seq m_2/m_1 \slt 1$, one has
$\eta \seq q/(1+q)^2$ and $q\seq (2\eta)^{-1} [1-(1-4\eta)^{1/2}] -1$,
which becomes $ q \ssimeq \eta$ at $\eta \sll 1$.}
This gives a maximum of $\Vk \seq 175\kms$ for $\eta=0.19$,
corresponding to a mass ratio $q \seq 0.37$. The peak is rather narrow,
dropping to below $100\kms$ at $\eta \seq 0.24\sdash 0.25$ ($q \sgt 0.67$)
and at $\eta \slt 0.12$ ($q \slt 0.16$).
%
The dependence on $q$ points to a potential bottleneck that may suppress 
the initial growth of the SMBH during the first few mergers, 
when the SMBH is still of a low mass such that the mass ratio tends to be 
rather large, $q \sgt 0.1$. 
We learn that with no BH spins, or with highly aligned spins and orbit, 
this bottleneck can be overcome once the escape velocity is modest,
of order $100\sdash 200 \kms$, as expected in typical FFB galaxies.
However, we learn from the figure that this bottleneck can be more difficult 
to overcome for large spins and for misaligned spin and orbit unless the escape 
velocity is higher than expected for typical FFB galaxies. 

\subsection{Escape Velocity}
 
The recoil pushes the merged SMBH from the galaxy center to a maximum distance 
$\rhalt$, which one may wish to estimate as a function of $\Vk$.
Consider a mass profile $M(r)$ and approximate the circular velocity
as $\Vc^2 \seq G M(r)/r$ and the inward acceleration as
$a \seq G M(r)/r^2$ such that $a \seq \Vc^2/r$.
Assume that the inner rotation curve is rising with $r$ while the magnitude
of the inward acceleration is constant or declining with $r$.
This is the case, for example, for a self-gravitating exponential disk inside
its maximum velocity at $1.063\,\rexp$, or for an NFW DM halo inside its
radius of maximum velocity.
Given that at any $r \leq \rhalt$ along the way $a(r) \sgeq a(\rhalt)$,
we can then write
\be
\rhalt \leq \frac{\Vk^2}{2 a(\rhalt)} \, ,
\label{eq:rhalt}
\ee
where the r.h.s. is the halt radius for an initial velocity $\Vk$ and
a constant acceleration $a(\rhalt)$.
Using $a(\rhalt) \seq \Vc(\rhalt)^2/\rhalt$, we obtain
\be
\Vc(\rhalt) < \frac{1}{\sqrt{2}} \Vk \, .
\ee
This implies that as long as $\Vk \slt \sqrt{2} \Vmax$, one has 
$\Vc(\rhalt) \slt \Vmax$, such that $\rhalt \slt \rmax$, where
$\rmax$ is the radius where the rotation curve is at a maximum value $\Vmax$,
a characteristic radius for the galaxy.
This limits the recoil of the central BH to inside $\rmax$, where it can 
migrate back to the center on a timescale that is comparable 
to the migration timescale within the inner body of the disk
as estimated in \se{DF_disk}, provided that the recoil is in the disk plane.
We therefore consider the effective escape velocity to be 
$\Vesc \seq \sqrt{2}\Vmax$.
A recoil with a higher velocity will prevent this BH from growing into
a central supermassive BH.

In an exponential disk $\rmax \ssimeq 1.8 \rexp$ with
$\Vmax \ssimeq 1.063\Vexp$, where $\Vexp$ is the rotation velocity at the
exponential radius $\rexp$.
For an exponential disk at the FFB threshold line 
$\Mveight \seq (1+z)_{10}^{-6.2}$,
using the disk properties from \equ{Re} to \equ{exp_mass},
we obtain
\be
\Vc(\rexp) = 114 \kms\, \epsilon_{0.2}^{1/2} (1+\zffb)_{10}^{-1.6} \, .
\label{eq:Vexp_ffb}
\ee
This implies $\Vmax \seq 121\kms \epsilon_{0.2}^{1/2} (1+\zffb)_{10}^{-1.6}$,
and
\be
\Vesc = \sqrt{2}\Vmax = 171\kms \epsilon_{0.2}^{1/2} (1+\zffb)_{10}^{-1.6} \, .
\label{eq:Vesc}
\ee
This escape velocity for a fiducial FFB disk at $\zffb \ssim 10$
is rather similar to the maximum kick velocity for non-spinning merging BHs
in \equ{kick}.
It implies that, in the absence of BH spin, or when the spins are aligned
with the orbital angular momentum, the SMBH 
is not pushed to outside the central regions of the disk,
and it can be pulled back by dynamical friction to the galaxy center within the 
migration timescale estimated in \se{DF_disk}.
However, as indicated in \fig{spin}, spin-orbits misalignment introduces 
a non-trivial bottleneck which would require a higher escape velocity.

\begin{figure} 
\centering
\includegraphics[width=0.49\textwidth,trim={1.5cm 5.5cm 1.5cm 4.5cm},clip]
{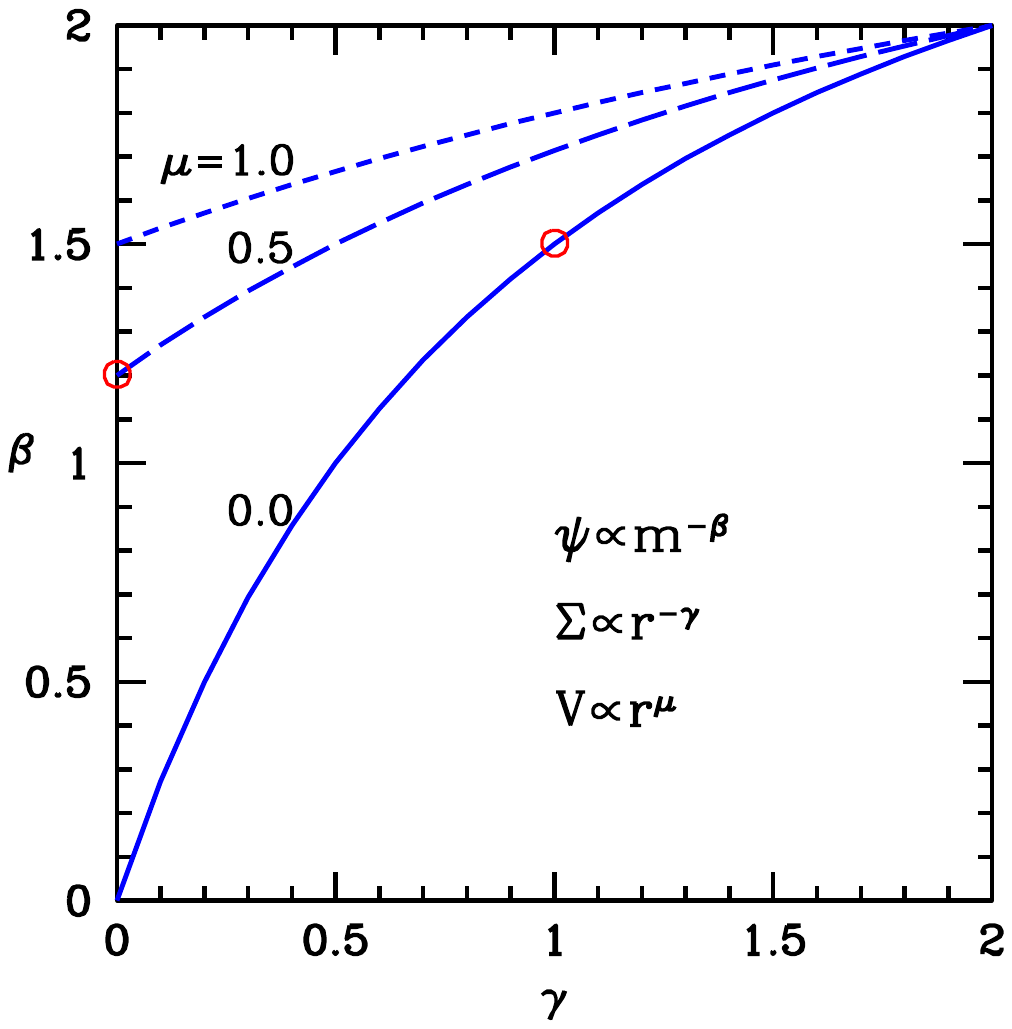}
\vspace{-15pt}
\caption{
The negative slope $\beta$ of the mass function of seed black holes as they
merge with the central SMBH, $\psi \prop m^{-\beta}$.
It is derived using \equ{psi} assuming an original BH mass function
$\phi \prop m^{-2}$ everywhere in the disk,
a disk of BHs with a power-law surface density profile
$\Sigma \prop r^{-\gamma}$,
and a total circular velocity $\Vc \prop r^{\mu}$ (which may include the
additional contribution of a halo and a post-compaction central mass).
At time $t$ and mass $m$, these two power laws are relevant for the radial
region from which BHs of mass $m$ arrive to the center at time $t$,
as derived for a hot disk using $\tdf \seq 0.5 \Vc/\adf$ and $\adf$ from
\equ{adf_hot}.
Self-gravitating disks with the shown $\mu$ are marked by red circles.
We read, for example, that for a flat rotation curve ($\mu \seq 0$) and
for the BHs that arrive from the disk range where $\Sigma \prop r^{-1}$
(a self-gravitating Mestel disk)
the mass function at the center is $\psi \prop m^{-1.5}$.
These values of $\beta$ refer to the mass function used in the recoil
simulations summarized in \fig{recoil_results}.
}
\vspace{-10pt}
\label{fig:mf}
\end{figure}

An additional bulge or DM halo would tend to increase $\Vmax$ and $\Vesc$.
In particular, the wet compaction events discussed in \se{compaction},
which are common in massive galaxies at high redshifts,
would significantly enhance the escape velocity, typically to $400\kms$ 
and possibly up to $700\kms$ \citep{lapiner23}. 
Based on \fig{spin}, such escape velocities may be high enough to overcome
the bottleneck introduced by GW recoil even for spinning BHs 
in relatively hot disks.
This will be quantified below using Monte-Carlo simulations of sequential BH 
mergers.

\begin{figure*} 
\centering
\includegraphics[width=0.90\textwidth]
{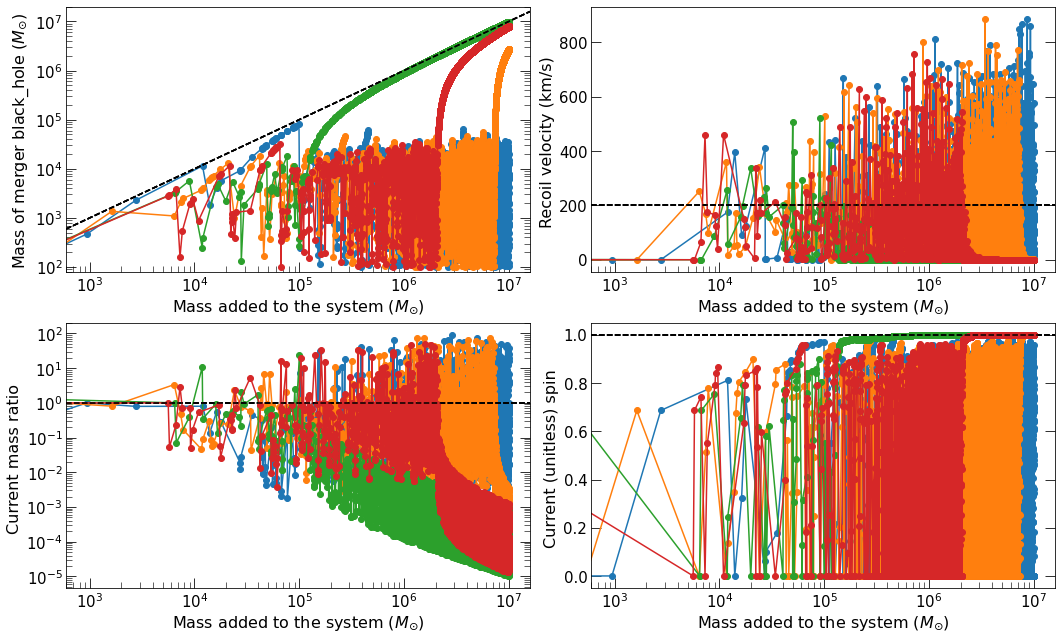}
\vspace{-5pt}
\caption{
SMBH growth by mergers in four simulations of one example case,
where the initial primary mass is drawn from the BH mass function of
$\beta \seq 1.13$, 
the spin-orbit misalignment angle is randomly drawn from $0-10^{\circ}$,
and $\Vesc \seq 200\kms$.
The sequence of mergers is represented along the x axis by the accumulated
mass of the merging BHs 
from the initial primary mass to a total of $10^7\msun$.
Top left: the mass of the merged BH after each merger.
Top right: the recoil velocity in comparison to the escape velocity.
Bottom left: the mass ratio (when it is larger than unity, the primary and
secondary exchange roles).
Bottom right: the spin of the primary BH.
When $\Vk \sgt \Vesc$ the SMBH is ejected and a new primary is chosen.
This is more likely to happen in the early mergers, when the mass ratio tends
to be high.
Three of the SMBHs manage to eventually grow while one fails (blue).
Growth is practically guaranteed once the SMBH mass exceeds $10^5\msun$,
while the secondary merging BHs are limited to $\leq\!10^4\msun$ such that the
mass ratio drops below $0.1$.
}
\label{fig:recoil_sim}
\end{figure*}

\subsection{The mass function of merging BHs}
\label{sec:mf}

In order to simulate the growth of the SMBH by a sequence of binary
mergers of seed BHs, one needs to know the effective mass function of the 
incoming BHs as they arrive at the galaxy center at a given time $t$.
The initial mass function of the seed BHs is assumed to follow that of the
FFB star clusters, \equ{phi},
which we assume to be the same at all radii in the galaxy disk, namely
\be
\phi(m) = \frac{\dd N}{\dd m} 
\prop m^{-\alpha} \, , \quad \alpha \slsim 2.0 \, ,
\label{eq:phim}
\ee
normalized such that $\int \phi(m) \, \dd m \seq M$, with $M$ the total mass in
BHs.
At time $t$, the desired mass function of merging BHs at the galaxy center, 
$\psi(m,t)$, is modulated by the number of BHs that started their inward 
migration at $t\seq 0$ in a radius within the ring $(r,r\splus \dd r)$ 
and reach the center during a time interval $(t,t\splus \dd t)$. 
The radius of origin of these BHs at $t\seq 0$, $r(m,t)$, is to be derived
from the expression for the dynamical-friction time at $r$.
At $t\seq 0$, the number of BHs of mass in the range $(m,m\splus \dd m)$
in the ring $(r,r\splus \dd r)$ is
\be
\frac{\dd N}{\dd m}\dd m 
=\phi(m)\, \dd m\, \frac{2\pi \Sigma(r)\,r\,\dd r}{M} \, .
\label{eq:Nt0}
\ee  
At $t$, the number of BHs in the same mass interval 
that arrive at the center during $(t,\dd t)$ is
\be
\psi(m,t)\, \dd m\, \dd t = \frac{\dd N}{\dd m\, \dd t}\, \dd m\, \dd t
= \frac{\dd N}{\dd m\, \dd r}\, \frac{\pa r}{\pa t}\, \dd m\, \dd t \, .
\label{eq:Nt}
\ee
Here $m$ and $t$ are the independent variables, and 
the second equality represents a transformation of variables from $t$ to $r$ 
using the Jacobian $\pa r/\pa t$ at a given $m$.
Extracting $\dd N/(\dd m\, \dd r)$ at $t \seq 0$ from \equ{Nt0}, and inserting
it in \equ{Nt}, we finally obtain 
\be
\psi(m,t) = \phi(m)\, \frac{2\pi\, \Sigma[r(m,t)]\, r(m,t)}{M}\, 
\frac{\pa r(m,t)}{\pa t} \, .
\label{eq:psi}
\ee

The desired $r(m,t)$ can be obtained, e.g., using  
the DF acceleration $a_{\rm df,hot}$ of \equ{adf_hot}, 
with $t \seq \tdf \seq 0.5\Vc/\adf$.
The radius $r(m,t)$ is extracted by solving the equation
\be
G^2 \tilde{F}\, m\,t = \frac{\Vc(r)^3}{\Sigma'(r)\, ,}
\label{eq:rmt}
\ee
where $\tilde{F} \seq 16\,\pi\Fd\, (r/h)^2$ is a dimensionless constant 
and $\Sigma'(r) \seq {\dd}\Sigma/{\dd}r$.

As an example, we consider a power-law surface density profile for the disk,
$\Sigma \sprop r^{-\gamma}$ with $0 \sleq \gamma \sleq 2$,
and a power-law circular velocity profile 
$\Vc \sprop r^{\mu}$ with $-0.5 \sleq \mu \sleq 1$.
For a self-gravitating disk we expect $\mu \seq (1-\gamma)/2$, but we allow
the circular velocity to include the additional contribution of a spherical
halo and a post-compaction central mass.
The solution of \equ{rmt} is $r \sprop (m\,t)^{1/(3\mu+\gamma+1)}$,
which gives in \equ{psi}
\be
\psi(m) \prop m^{-\beta} \prop \phi(m)\, m^{(2-\gamma)/(3\mu+\gamma+1)} \, ,
\label{eq:mf}
\ee
where $ \beta \seq \alpha -(2-\gamma)/(3\mu+\gamma+1)$.
\Fig{mf} shows the negative slope $\beta$ as a function of 
$\gamma$ and $\mu$, for an assumed
$\alpha \seq 2$ in \equ{phim}.
For a BH mass $m$ at time $t$, 
one can interpret the power-law profiles with $\gamma$ and $\mu$ as 
local approximations near $r(m,t)$.
After a certain time, when the massive BHs arriving at the center 
have originated from the main body of the disk,  
one may expect $\gamma \ssim 1$ or even steeper.
For a self-gravitating disk, starting with $\alpha \seq 2$,
we obtain $\beta \seq 1.2,1.5,2.0$ for the cases
$(\gamma,\mu) \seq (0,0.5), (1,0), (2,-0.5)$ respectively.
We adopt as our fiducial case $\psi \sprop m^{-1.5}$,
as derived for a self-gravitating Mestel disk, $(\gamma,\mu) \seq (1,0)$.

As another example, 
for a self-gravitating exponential disk with a scale radius $r_1$, 
at early times when $r \sll r_1$, \equ{rmt} gives $r \sprop m^{2/3}$,
which yields in \equ{psi} $\beta \seq 0.67$.

We assume as our crude fiducial slope for the seed BHs $\alpha \seq 2$, 
only slightly steeper than the original $\alpha \seq 1.8$ for the clusters
from simulations.
The effect of stellar capture within the clusters, based on \equ{delta},
could in principle lead to a steeper slope at low masses and late times, 
but considering the disruption of clusters over time, we conservatively
adopt only a modest steepening to $\alpha \seq 2$.

\subsection{Monte Carlo simulations of Recoils}

Using fitting formulae for the recoil velocity as a function of the
given merger parameters, namely the mass ratio, the BH individual spins 
and the spin-orbit alignments \citep[based on][]{gerosa16,gerosa23}, 
we ran Monte Carlo 
simulations of SMBHs as they grow by a sequence of binary mergers. 
We wish to evaluate the fraction of galaxies within which the SMBH would 
overcome the early recoil bottleneck as a function of 
(i) the escape velocity, 
(ii) the thickness of the disk of BH seeds,
which determines the spin-orbit alignment and SMBH spin buildup, 
(iii) the choice of the initial primary central BH mass, 
and (iv) the mass function of BHs as they merge with the SMBH
via the slope $\beta$ of \se{mf}. 

In each sequence,
the potential SMBH starts from a given primary BH of mass $m_0$. 
Its mass is either drawn at random from the mass function of seed BHs, 
or it is fixed at an assumed value, e.g., $m_0 \seq 10^4\msun$ or $10^5\msun$. 
The latter scenario may be more realistic if a different high-$z$ process, 
such as direct collapse of a mini-halo \citep{inayoshi20}, 
produced a single high mass BH seed that reached the galaxy center before the 
BHs produced in FFB clusters could arrive there.

\begin{figure} 
\centering
\includegraphics[width=0.49\textwidth] 
{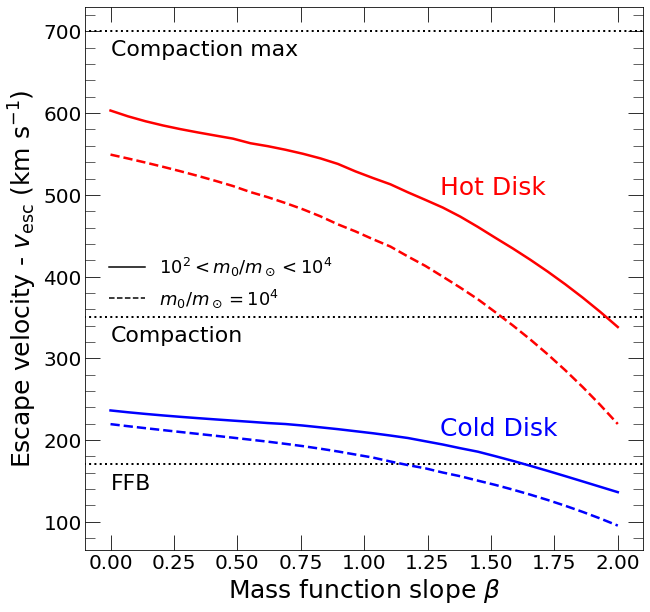}
\vspace{-5pt}
\caption{
Simulation results for the the GW recoil bottleneck.
Shown is the threshold escape velocity $\Vesc$ that is required for SMBH growth
in at least one third of the cases as a function of the slope $\beta$ of the BH
mass-function
$\psi \sprop m^{-\beta}$.
The blue lower curves are for cold disks, with spin-orbit misalignments
$\theta \seq 0 \sdash 10^\circ$, while
the red upper curves correspond to hot disks, with
$\theta \seq 0 \sdash 30^\circ$.
The solid and dashed curves refer to cases where the mass of the initial
primary BH $m_0$ is either drawn from the seed BH mass function in
$(10^2 \sdash 10^4\msun)$ or selected to be $10^4\msun$, respectively.
The fiducial escape velocity for an FFB disk at $\zffb \seq 9$ is
$\Vesc \ssimeq 170\kms$ when ignoring compaction events,
and it could rise to several hundred $\kms$ after compaction
(horizontal dotted lines).
Naturally, the required escape velocity for BH growth is lower for a
steeper BH mass function.
For no BH spins, or fully aligned spin and orbit,
there is always growth for $\Vesc \sgeq 100\kms$
(not shown).
For a cold disk, The fiducial $\Vesc \ssim 200\kms$ of FFB is sufficient
for BH growth if $\beta \sgeq 1.5$ (or $\beta \sgeq 1$) for
a random $m_0$ (or $m_0 \seq 10^4\msun$).
The bottleneck becomes more severe for a hot disk,
where a non-negligible growing fraction can be obtained either
with a $\Vesc$ that has been boosted by compaction events to the level of a few
hundred $\kms$, or with the original FFB $\Vesc \ssim 200\kms$
if $m_0 \sgt 3\stimes 10^4\msun$ (not shown).
}
\vspace{-10pt}
\label{fig:recoil_results}
\end{figure}

The evolution of each galaxy, with a given escape velocity, is represented by 
a sequence of binary mergers, for which the mass of the secondary is drawn 
at random from the mass function in \equ{mf} with a given value of $\beta$.
The simulation continues till the total accreted mass amounts to $10^7\msun$,
motivated by the SMBH masses obtained in \se{DF}
and detected at $z \seq 4 \sdash 7$.
The post-merger central BH mass, calculated using \citet{barausse12}, 
becomes the new primary mass for the subsequent merger unless 
$\Vk \sgeq \Vesc$, in which case the
merged remnant is ejected, and the sequence restarts with a new primary seed
of mass $m_0$.  
The post-merger central BH spin is assumed to be the sum of the spins of the 
two merging BHs 
and the angular momentum of the orbit with the appropriate numerical-relativity 
corrections \citep{hofmann16}. For this purpose we approximate the spins to 
be aligned with the galactic angular momentum as the misalignment angles 
are always small. 
The spin-orbit angles are drawn at random in the range 
$0 - \theta_{\rm max}$, with $\theta_{\rm max} \seq 0, 10^\circ, 30^\circ$. 
These three choices represent pure alignment, a cold disk and a hot disk, 
respectively. We take all spin-orbit misalignments to be prograde, 
assuming that the orbit and the spins 
reflect the same original galactic disk angular momentum, and that the seed 
BHs are confined to the disk plane more than the stars 
due to dynamical friction in the vertical direction \citep{stewart00}.

\Fig{recoil_sim} shows some details of four example Monte-Carlo 
simulations of mergers of seed BHs. 
Here $\theta_{\rm max} \seq 10^\circ$ and the initial primary BH drawn at 
random from the same seed mass function (in this case with $\beta \seq 1.13$).
One can see many repeated ejections of the SMBH, where the merged BH mass
drops sharply to a new $m_0$. These occur preferentially in the first few 
mergers of each sequence, where the BH mass ratios are high. 
Nevertheless, despite the high cumulated spins, 
three out of the four cases eventually 
reach the continuous growth phase in which mergers of increasingly diminished 
mass ratios yield lower and lower recoil velocities.

\Fig{recoil_results} presents a summary of our Monte-Carlo simulation results 
for the recoil bottleneck.
We show the threshold escape velocity, $\Vesc$, that is required for SMBH 
growth (exceeding $10^6\msun$) in at least one third of the simulated cases
out of 1,000 cases in each of our experiments,
as a function of the slope $\beta$ of the mass-function of the BHs
as they reach the galaxy center.
The fraction of growing SMBHs rises steeply from zero to unity 
as a function of $\Vesc$ near the critical value shown. 
The different curves refer to different levels of spin-orbit misalignments, 
resulting from the `cold' and `hot' disks specified above,
and to two different choices of the primary central BH at the start of each
sequence, both rather conservative choices.
Recall that
the fiducial escape velocity for an FFB exponential disk at $\zffb \seq 9$ is
$\Vesc \ssimeq 170\kms$ when ignoring the effects of compaction events,
and it could rise to several hundred $\kms$ after compaction.

Not shown in the figure is that  
we find 100\% SMBH growth for the cases of aligned spin-orbit or no spins,
even for a random initial primary mass, for any escape velocity larger than
$100\kms$
and all choices of $m_0$ and $\beta$ in the ranges considered.
This is despite possible failures during the first few mergers in certain cases.
We also find 100\% growth for more massive central primaries of
$m_0 \seq 10^5\msun$ in all the cases tested for
$\beta$, $\theta_{\rm max}$ and $\Vesc \sgt 100\kms$.

We see in \fig{recoil_results} that,
in general, for a given $\Vesc$, 
the growing fraction becomes larger for steeper BH mass functions, 
as they favor lower mass ratios that cause weaker recoils. 
For a steep enough mass function ($\beta \sgeq 1.5$ for a random primary mass, 
and
$\beta \sgeq 1$ for a $10^4\msun$ primary), significant SMBH growth is allowed 
for the fiducial escape velocity of $\sim\! 200 \kms$ as determined at the FFB
phase for an exponential disk ignoring subsequent compaction events.

The bottleneck becomes more severe for a hot disk
($\theta \seq 0 \sdash 30^\circ$). 
Here, with $\beta \seq 1.5$, and for $m_0 \seq 10^4\msun$,
the SMBH growing fraction becomes non-negligible only for $\Vesc \sgt 350\kms$.
We find (not shown) that this threshold drops to $\sim\!200\kms$ once the 
primary exceeds $3\stimes 10^4\msun$.
We learn that in order to overcome the GW recoil bottleneck in a hot disk 
with a $10^4\msun$ primary central BH
the escape velocity should be boosted to the level of a few hundred $\kms$, 
possibly by the compaction events that are expected to be common at the
relevant redshifts.


\adr{
A caveat of the current toy-model analysis is that it considers SMBH growth 
through a sequence of binary mergers, while the large number of BHs is likely 
to also lead to multiple merger chains, resulting in the growth of more than 
one SMBH. In such cases, large-recoil mergers of mass ratios $>\! 0.1$ 
are more likely, and the associated displacements of the central BH from the 
high-density core region would suppress its interaction with other sinking BHs 
and thus complicate the buildup of a single SMBH. This should be studied by 
non-trivial simulations that incorporate a proper treatment of multiple BH 
mergers.
}

\section{Compaction-Driven Black-Hole Growth}
\label{sec:compaction}

The SMBH growth is likely to be boosted by galactic wet compaction events,
common at high redshifts,
which affect both the BH inward migration rate estimated in \se{DF} and the
results of GW recoil discussed in \se{recoil}.

The process of `wet compaction' to a `blue nugget' was introduced in
\citet{db14} and \citet{zolotov15}. The process and its far-reaching
implications on all major galaxy properties were studied in these and in
a series of subsequent papers, largely based on the VELA zoom-in cosmological
simulations, and they are reviewed in \citet{lapiner23}.
According to the simulations, this is a generic process that occurred in the
history of most galaxies, preferably when the DM halo reaches a ``golden mass"
of $\sim\!10^{11.5}\Msun$. Triggered by drastic angular-momentum loss, e.g.,
due to wet galaxy mergers or collisions of counter-rotating inflowing streams,
gas is pushed to the central region of the galaxy, where it causes a starburst.
This results in a cuspy gas-rich `blue nugget' that passively evolves to a
compact stellar `red nugget'. This central mass concentration allows the
formation of an extended gaseous disk by stabilizing it against inflow
due to violent disk instability \citep{dekel20_ring}.
A similar picture, where the galaxies develop a baryon-dominated center
once above a threshold stellar mass $\sim\!10^9\msun$ at all redshifts
in the range $z\seq 0\sdash 6$, is confirmed based on the TNG50 simulations
\citep{degraaff24}.

\citet{lapiner21}, using the NewHorizon cosmological simulations which
incorporated black holes,
discussed the process of compaction-driven BH growth above the golden mass.
They noticed that in pre-compaction galaxies of lower masses,
the BHs tend to wander about the center, while post compaction they become
confined to the galactic centers. This is interpreted as a result of the
compaction-driven deepening of the central potential well and the enhancement
of dynamical friction exerted on the BH by the dense central baryons.

The VELA, NewHorizon and TNG50 simulations allow a study of compaction events 
at $z \ssim 6$ and later; this specific redshift range may be an artifact 
of the low abundance of massive enough galaxies at higher redshifts in these 
simulations.  
In reality, compaction events
might have occurred above a similar threshold mass at higher redshifts as well.
We estimate that the typical FFB galaxies, that form at $\zffb \ssim 9$,
indeed reach the threshold mass for compaction by $z \ssim 7$, since a halo of
$\ssim 10^{11}\msun$ at $\zffb \ssim 9$ is expected on average to grow
by $\sim\!1$ dex until $z \ssim 6$
\citep[assuming $M \prop e^{-0.8 z}$][]{dekel13}.

The compaction events can serve the SMBH growth in several ways.
First,
in cases where the disk-plus-halo density profile and the associated
$\Omega(r)$ profile are too flat for efficient inward migration by dynamical 
friction, as discussed in \se{DF},
the steepening of these profiles by the wet compaction events may be
necessary for ensuring continuing inward migration all the way to the galactic
center, avoiding core stalling and possible buoyancy.
Indeed, as can be seen in Fig.~24 of \citet{lapiner23},
the post-compaction angular-velocity profile tends to be declining steeper than
$\Omega(r) \sprop r^{-1}$ at all radii, typically enough for avoiding core
stalling \citep{read06,goerdt10,kaur18,dutta19,banik21}.

A second crucial role of the compaction events concerns overcoming the
bottleneck introduced by the strong GW recoils of the SMBHs
that result from misaligned spins and orbit of the merging BHs, as
discussed in \se{recoil}. 
By increasing the escape velocities from the galaxy centers,
the compaction events may be important for breaking through this bottleneck, 
which can be non-trivial in the case of thick galactic disks.
In the VELA zoom-in cosmological simulations, 
as seen in Fig.~B1 of \citet{lapiner23},
the post-compaction circular velocity within the effective radius
$\Re \ssim 1\kpc$ is typically $250\kms$, and it ranges from $160\kms$ 
to $500\kms$.
Similarly, 
in the NewHorizon simulated galaxies with SMBHs, as seen in Fig.~4 of
\citet{lapiner21}, the post-compaction circular velocity at $1\kpc$
is typically $300\kms$, and it ranges from $120\kms$ to $500\kms$.
We can thus consider a typical post-compaction escape velocity of
$\Vesc \ssim 350\kms$, ranging from $170$ to $700\kms$.
As found in \se{recoil},
at the high end this is enough for keeping the SMBH at the galaxy center
even in cases of misaligned spins and orbit in relatively thick disks. 

Third, as discussed in \se{final_pc},
the wet compaction events can help solving the final parsec problem if 
triaxiality and three-body mergers fail to do so for a stellar system.
The gas pushed into the galaxy center feeds an AGN accretion disk
that can exert compressive circum-binary torques \citep{armitage02}. 

As a word of caution we note that major galaxy mergers, one kind of the 
possible drivers of wet compaction events, may induce significant global 
morphological and kinematical changes in the structure of the galaxies that 
host the migrating BHs, and possibly add a mechanism of BH ejection.
These effects are yet to be investigated.
\adr{
Another caveat is associated with the tendency of compaction events to occur
near a `golden mass' of $\sim\! 10^{10}\msun$ \citep{dekel19_gold,lapiner23}.
This introduces a potential difficulty for galaxies of much lower masses to 
retain post-merger over-massive BHs, in potential conflict with some of
the low-mass BH hosts observed using {\tt JWST}. For these low-mass galaxies 
to grow over-massive BHs by a merger-driven scenario, one has to appeal to the 
less frequent compaction events that occur below the golden mass.
}
\adr{
A third caveat associated with compaction events in the BH context is that 
they tend to lead to starbursts that increase the stellar mass
\citep{tacchella16_ms} and thus make it more difficult to obtain the 
required high values of $\fbh$.  This will have to be quantified 
using simulations that incorporate compaction events during the process of BH 
migration and mergers.
}

\begin{figure*} 
\centering
\includegraphics[width=0.99\textwidth,trim={0.0cm 1.8cm 0.0cm 2.2cm},clip]
{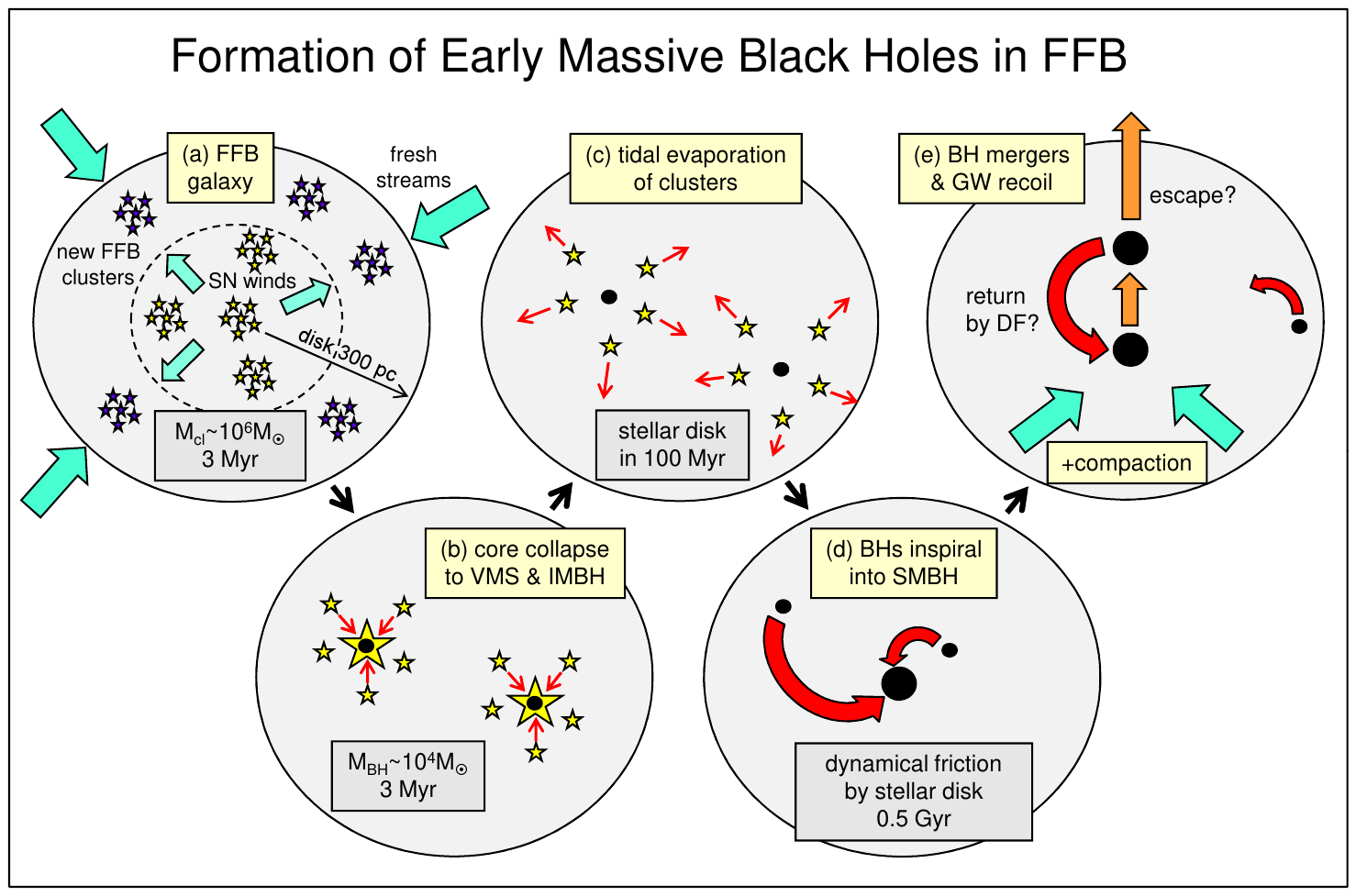}
\vspace{-5pt}
\caption{
A cartoon summarizing the scenario of merger-driven BH growth in FFB
galaxies at cosmic dawn.
(a) Feedback-free starbursts in thousands of clusters at $z \ssim 10$,
each forming in less than $3\Myr$ with a typical mass $\sim\!10^6\msun$,
in $\sim\!10$ generations of $\sim\!10\Myr$ each
over a global FFB period of $\sim\!100\Myr$.
(b) Rapid core collapse in each cluster during $\sim\!3\Myr$, sped up by the
presence of massive stars and by cluster rotation and flattening, leading to a
very massive central star (VMS) that becomes an intermediate-mass BH (IMBH) of
$\sim\!10^4\msun$.
(c) Tidal evaporation of the clusters, progressing inside-out in the galactic
disk over $\sim\!100\Myr$, producing a background stellar disk capable of
exerting effective dynamical friction.
(d) dynamical-friction-driven inspiral of the BHs that accumulate at the
galactic center total BH mass of $\sim\!10^{7\pm 1}\msun$ in $\sim\!0.5\Gyr$,
namely by $z \ssim 7$.
(e) SMBH growth by BH mergers, subject to GW recoils during the first stages of
growth. The SMBH is retained if the galactic disk of BHs is cold or if the
potential well is deepened by wet compaction events.
}
\vspace{-10pt}
\label{fig:bh_cartoon}
\end{figure*}

\section{Conclusion}
\label{sec:conc}

We verified the conditions under which the scenario of feedback-free 
starbursts at cosmic dawn \citep{dekel23,li24}
may provide a natural setting for the formation of 
intermediate-mass seed black holes and their subsequent growth, largely
merger-driven, to super-massive black holes with a high BH-to-stellar mass 
ratio as indicated by JWST observations
\citep{pacucci23}.

\Fig{bh_cartoon} is a cartoon summarizing the main stages of the
BH growth scenario that has been examined here. 
It starts with thousands of star clusters formed during the FFB phase 
within which rapid core collapse leads to intermediate-mass BH seeds.  
The FFB phase is followed by cluster disruption which causes the efficient 
inspiral of the BHs into the galactic center, where the BHs merge into a 
SMBH that may overcome GW recoil provided that the disk is cold or the 
potential well is deepened by wet compaction events. 

In the FFB scenario,
the sites for the formation of intermediate-mass seed black holes are the 
thousands of star clusters that serve as the building blocks of the FFB 
galaxies,
each starbursting in a free-fall time of a few Myr during the lifetime of their
massive stars and before the onset of stellar and supernova feedback.
The seed black holes form by rapid core collapse in the FFB clusters
\citep{lynden68} on a similar free-fall timescale, sped up by their young, 
broad stellar mass function. This process is further hastened by their internal
rotation and associated spatial flattening in the disk version of the FFB 
scenario.
The core collapse is driven by the inward migration of
the short-lived massive stars due to the mass segregation by two-body
interactions and dynamical friction,
and it is sped up by the spatial flattening and the gravo-gyro instability 
\citep{hachisu79}
that are induced by the cluster internal rotation within the galactic disks
\citep{ceverino12}.

The FFB-generated seed BHs eventually migrate by dynamical friction to the 
centers of the compact galactic disks, ready to merge into SMBHs with high 
BH-to-stellar mass ratios.
The dynamical friction is exerted mostly by the compact stellar 
galactic system that forms by the tidal disruption of the FFB clusters.
The compact disk morphology in the disk version of the FFB scenario speeds up 
this dynamical friction process to the level required for matching the SMBH 
masses and high $\fbh$ ratios that are indicated by observations at 
$z \ssim 4\sdash 7$ as well as at $z \ssim 1 \sdash 3$.

The SMBH growth by BH mergers has then to overcome the bottleneck introduced by
GW recoil velocities that may exceed the escape velocity from
the galaxy \citep{pretorius05,campanelli06,baker06}.
These recoil velocities can be large, especially during the first mergers where 
the BH mass ratio can be not much smaller than unity, and particularly so for 
large spin-orbit misalignments. A large fraction of growing SMBHs is 
nevertheless obtained for relatively cold disks of seed BHs. 

The SMBH growth can be boosted up in three different ways by 
the generic wet compaction events that tend to occur at high redshifts above
a threshold galaxy mass \citep{zolotov15,lapiner23}.
First, the compaction-induced cuspy density profiles with steeply declining 
angular-velocity profiles could help avoiding core stalling in the inward 
migration of BHs within the inner galactic disks.
Second, the compaction events help overcoming the GW recoil bottleneck by 
increasing the central escape velocities from the galaxies.
Third, the push of gas into the galaxy center can help enabling the
final parsec approach of the merging BHs.

Our more quantitative conclusions are as follows.

\no\bul 
For a standard IMF in rotating, flattened FFB clusters at $z \ssim 10$,
BH seeds of $\sim\!10^4\msun$ (and smaller) are expected to form by core 
collapse in clusters of $\sim\!10^6\msun$ (and below),
giving rise to particularly high BH-to-stellar mass ratios of $\fbh \ssim 0.01$.

\no\bul 
Such BHs are expected in most of the FFB clusters but not in the
most massive clusters near the Toomre mass of $\sim\!10^7\msun$. 
However, with a very top-heavy IMF in disky, strongly rotating
FFB clusters the core collapse could possibly be sped up to form 
$\sim\!10^5\msun$ BH seeds in the $\sim\!10^7\msun$ clusters.

\no\bul 
Based on our estimate of the dynamical friction rate in an FFB compact galactic 
disk, most of the BH seeds of $10^4\msun$ should migrate to the galaxy center
and provide mass for a SMBH of $\sim\!10^{7\pm1}\msun$ by $z \ssim 4\sdash 7$. 
Central SMBHs are thus capable of inheriting the high $\fbh$ ratio of 
$\sim\! 0.01$, which originates in the process of runaway intermediate-mass BH 
formation in the FFB clusters, and matches observational estimates.

\no\bul 
A significant fraction of the SMBHs can overcome the bottleneck
introduced by GW recoils and grow by mergers 
once the seed BHs inspiral within a relatively cold galactic disk such that
the spin-orbit misalignments are limited to $0 \sdash 10^\circ$ and the 
escape velocity is near the FFB fiducial value of $\sim\!170\kms$ or larger.

\no\bul 
Wet compaction events are expected in galaxies once they grow above a threshold
halo mass of $\Mv \ssim 10^{11.5}\msun$ \citep{lapiner23}.
This steepens the decline of the central density profile and angular-velocity
profile and thus suppresses potential core stalling of the inward migration of 
the BH seeds.

\no\bul 
These compaction events typically increase the escape velocities to several 
hundred $\kms$ and thus enable SMBH growth also in somewhat hotter disks of 
BH seeds, with spin-orbit misalignments of $\sim \!30^\circ$. 
In addition, the supply of central gas can help overcoming the final parsec 
problem in the BH mergers.

One should be aware that our current analysis is to be be interpreted with 
caution and serve primarily as a feasibility study  
because it is based on simplified modeling and involves large uncertainties.
For example, in our idealized model
the migrating BH seeds are assumed to coalesce into the SMBH, largely ignoring
further growth of the BH by accretion of gas and stars. 
The seed BHs are assumed to keep their masses during the migration,
ignoring binary BH mergers before coalescence in the center.
The galaxy is very crudely assumed to remain static in time since the FFB phase
ignoring galactic accretion and star formation at later times except for 
crudely considering the qualitative effects of wet compaction events.

There are rather surprising significant uncertainties in the basic 
physical processes that are involved in our crude modeling. 
One such uncertainty is associated with the gravo-thermal-gyro core-collapse 
time and the resultant intermediate-mass BH-seed mass within rotating, 
flattened clusters.  The unknown width of the IMF adds to this uncertainty.
A second severe uncertainty concerns the strength of
dynamical friction in a disk, which governs 
the assembly to a SMBHs is the compact FFB galactic disks,
and possibly also the core collapse in rotating disky clusters.
Beyond their use in the current study,
these two fundamental processes deserve detailed reliable numerical studies
using N-body simulations that are currently missing from the literature.

The poorly constrained galactic disk morphology in the FFB phase
introduces another uncertainty in the inward migration timescale, which
may involve core stalling. In such cases, wet compaction events are necessary
for allowing the completion of the migration process.
Finally, the unknown flattening of the disk of inspiraling BHs 
and the wide range of possible escape velocities introduce an uncertainty 
in the ability of the merged SMBHs to survive
GW recoils during the first mergers, as long as the SMBH has not grown
significantly yet.

We considered here the minimum growth path of SMBHs by BH mergers, 
ignoring further SMBH growth by gas accretion. If the growth by mergers indeed
dominates at high redshifts, it would evade the low-redshift \citet{soltan82} 
relation. This relation is the observed consistency between the quasar 
luminosity density and one tenth of the SMBH mass energy density, which 
indicates that most SMBHs acquired the bulk of their mass through radiatively
efficient AGN episodes associated with gas accretion.
If the high-redshift SMBH growth is indeed mostly by mergers, 
the ratio between the AGN luminosity function and the SMBH volume density is
expected to be lower than in lower-redshift quasars, and the SMBHs are expected
to be of low X-ray luminosities, as observed. 
Furthermore, the AGN feedback is likely to be less effective, not intervening
in the otherwise higher efficiencies of star formation predicted at high
redshifts.

Our successful current feasibility study motivates a more quantitative study  
of the proposed scenario of mostly merger-driven BH growth starting with the 
FFB phase at cosmic dawn.


\medskip
Data and results underlying
this article will be shared upon reasonable request to the corresponding author.

\begin{acknowledgements}
We are grateful for stimulating discussions with Martin Haehnelt, 
Thorsten Naab, Christophe Pichon, Priya Natarajan, Re'em Sari, Volker Springel
and Marta Volonteri. 
This work was supported by the Israel Science Foundation Grants
ISF 861/20 (AD, ZL), 2565/19 (NCS), 2414/23 (NCS), and 3061/21 (NM, ZL),
by the Bi-national Science Foundation grants 2019772 (NCS), 2020397 (NCS),
and 2020302 (NM),
by the NSF-BSF grant 2023730 (AD),
by the National Science Foundation grants AST-2307280 (FvdB)
and AST-2407063 (FvdB),
by an IASH postdoctoral fellowship (DDC),
and by a Horizon-MSCA postdoctoral grant 101109759 (ZL).
\end{acknowledgements}



\bibliographystyle{mnras} 
\bibliography{z10}

\end{document}